\documentclass[apj,twocolumn]{openjournal}
\usepackage{tikz}
\usetikzlibrary{arrows.meta,decorations.pathmorphing,calc,positioning}
\usepackage{newtxtext,newtxmath,enumitem,multirow,ctable}
\usepackage[breaklinks,colorlinks,citecolor=blue,urlcolor=blue]{hyperref}
\usepackage{comment}
\usepackage{orcidlink}
\usepackage{makecell}
\usepackage[caption=false]{subfig}

\newcommand{\kms}{km~s$^{-1}\,$}

\shorttitle{Dependences of radio pulsar parameters on the kick velocity}
\shortauthors{Lazarev \& Popov}

\begin{document}

\title{Dependences of radio pulsar parameters on the kick velocity\vspace{-3em}}


\author{Anton~Lazarev\altaffilmark{1}\orcidlink{0009-0004-9016-7812}}
\altaffiltext{1}{These authors contributed equally to this work.}
\email{anton.d.lazarev@gmail.com}
\affiliation{Department of Physics, Lomonosov Moscow State University, Leninskie Gory 1, Moscow, 119991, Russia}

\author{Sergei~Popov\altaffilmark{1}\orcidlink{0000-0002-4292-8638}}
\email{sergepolar@gmail.com}
\affiliation{Sternberg Astronomical Institute, Lomonosov Moscow State University, Universitetsky pr. 13, Moscow, 119234, Russia}

\begin{abstract}
According to several studies, analysis of observational data and theoretical modeling favor a bimodal distribution of the natal velocity kick of neutron stars.
We analyze this proposal by using available data on radio pulsars.
For $\sim200$ normal isolated radio pulsars with well-measured spin and kinematic parameters, we determine if they belong to the low- or high-velocity mode of such a distribution by applying the parametrization proposed by Igoshev (2020). Our results demonstrate that about $23\%$ belong to the low-velocity mode. We then analyze the differences in the properties of the two sets of pulsars belonging to the two modes. For some parameters (characteristic ages and distances), we see a clear difference between the two modes. However, for these quantities, it can be easily attributed to selection bias. For those parameters that are not a subject of strong selection, such as pulse width, we do not observe any difference. Interestingly, we detect a notable difference in the magnetic field distribution between the two modes. Lower field pulsars ($B\lesssim 10^{12}$~G) are overabundant among objects from the low-velocity mode in comparison to the high-velocity one. 
In particular, among pulsars with low fields ($\lesssim  10^{11}$~G), we do not identify any objects from the high-velocity mode of the kick distribution.
The origin of this discrepancy is not clear, and we discuss several possibilities.
Our analysis demonstrates that most probably, this feature can be explained by selection effects. 
In particular, the difference in the magnetic field distribution for low- and high-velocity pulsars demonstrates complicated behavior when an upper limit on the characteristic age of pulsars is introduced.
Thus, we conclude that there is no robust bimodality in physical parameters of radio pulsars that can be related to the proposed bimodality of the kick velocity. This can be considered as an indirect argument against the hypothetical bimodality
in the NS kick distribution.
\end{abstract}



\maketitle

\section{Introduction}
\label{sect:intro}

 Soon after the discovery of radio pulsars, it was noted that their spatial velocities substantially exceeded the velocities of their progenitors. Thus, a discussion of the origin of this discrepancy was initiated \citep{GunnOstriker1970ApJ, 1970SvA....13..562S}. The standard view is that neutron stars (NSs) obtain an additional momentum (kick) due to supernova explosions, see a review in \cite{2025NewAR.10101734P}. However, alternative models, in which an NS is accelerated for a long period of time, are also proposed in the literature (e.g., \citet{2022ApJ...931..123L, 2023MNRAS.522.5879A} and references therein).

Various types of objects and approaches are used to probe the kick velocity distribution of NSs. E.g., recently \cite{2025ApJ...985...12W} used data on proper motions, radial velocities, and parallaxes of several dozen NS X-ray binaries to derive the kick velocity distribution applying population synthesis modeling. \cite{2026arXiv260315750C} used the proper motions of magnetars for their velocity determination. However, the majority of results on the kick velocity distribution were obtained by analyzing radio pulsar properties (see e.g., \citet{2025A&A...700A..75D} and references therein).
 
 Observations of radio pulsars typically provide 2D (transverse) velocities based on their proper motion and distance measurements (e.g.,  \cite{2024MNRAS.530..287S} and references therein). It is non-trivial to derive the natal kick distribution from the observed 2D velocities. The first reason is that the radial velocity component is not known from observations, so additional assumptions about its value are necessary. Another reason is related to the fact that the present-day spatial velocity depends not just on the kick (especially for small values), but also on the progenitor's velocity (including the orbital velocity if the pulsar was born in a binary system unbounded after the supernova explosion) and on kinematic evolution in the Galactic gravitational potential. Finally, it should be noted that velocity measurements are subject to selection effects. For example, at larger distances, it is difficult to measure small velocities; on the other hand, high-velocity pulsars rapidly move out of the volume of high sensitivity of radio surveys.\footnote{Additional selection effects can appear if kick velocity is correlated with some parameters important for the detectability of pulsars, e.g., magnetic field.} To summarize, even a sample of a few hundred pulsars with well-measured proper motions and distances does not allow for a precise and unique determination of the natal kick velocity distribution.
 
In half a century, many attempts have been made to derive the distribution of natal kicks \citep{2025NewAR.10101734P}. Some approaches are based on samples of young pulsars for which the effects of evolution in the Galactic gravitational potential are not that severe, while others are based on population synthesis studies. Starting with the paper by \cite{2002ApJ...568..289A}, the bimodal kick distributions began to attract much interest. 
In particular, a set of more and more detailed models was proposed by Igoshev and his co-authors \citep{2017A&A...608A..57V, 2020MNRAS.494.3663I, 2021MNRAS.508.3345I}. Analysis of binary systems 
also favors a bimodal kick velocity distribution, e.g.\citep{2025ApJ...985...12W}.
Note, however, that \cite{2025ApJ...989L...8D} argues against bimodality and demonstrate that the results by \cite{2017A&A...608A..57V} and \cite{2020MNRAS.494.3663I} can be explained by low-number statistics. \cite{2025ApJ...989L...8D} suggest that
the underlying kick velocity distribution has a unimodal lognormal shape.

Present-day pulsars' velocities are in the range from a few \kms up to several thousand \kms \citep{2013ApJ...765...36M}. Large velocities can significantly influence the fate of binary systems \citep{2014LRR....17....3P} and the evolution of isolated NSs \citep{2026JHEAp..5300643A}. Due to kicks, a fraction of NSs can become halo objects or even escape the Galaxy \citep{2010A&A...510A..23S}.
However, kick velocities are of great interest, not only because they determine the evolution and observational appearance of NSs. It is essential that the kick is an imprint of the supernova explosion, see recent reviews, e.g., in \cite{2024Ap&SS.369...80J, 2020LRCA....6....3M}. Thus, detailed knowledge of the kick properties can improve our understanding of the supernova mechanism. 
If, in addition, some correlations between the NS parameters and the kick velocity are clearly identified, this can further advance supernova models \citep{2024ApJ...964L..16B}.

In this study, our goal is not to derive another kick velocity distribution from the observed properties of pulsars.  
Instead, with a large sample of pulsars, we try to probe the hypothesis of the bimodal distribution in a specific way. Our objective is to identify the belonging of radio pulsars with well-measured spin and kinematic parameters to the low or high kick velocity mode. For this task, we choose the distribution proposed by \cite{2020MNRAS.494.3663I}. Using coordinates, distances, spin properties, and proper motions of pulsars, we calculate their natal velocities and separate them between the two modes using a simple statistical criterion.
Then, we analyze whether the pulsar parameters differ between the two modes and discuss the differences. We hypothesize that if the bimodal distribution is real and reflects variations in the supernova parameters, then some distinctions between cumulative properties of pulsars from the two modes must appear.
In the next section, we describe our approach. Then, in Sect.~\ref{sect:results}, we present our results. The results are discussed in Sect.~\ref{sect:discussion}. Finally, in Sect.~\ref{sect:conclusions} we present our conclusions.

\section{Model}
\label{sect:model}

In this section, we present our model, which is used to classify radio pulsars into low- or high-velocity modes of the distribution. 
 
Our goal is to estimate the natal kick velocity of radio pulsars. To achieve this goal, for each pulsar in the sample, we integrate the pulsar's trajectory back in time to its birthplace. In this procedure, several assumptions have to be made. First, we have to specify with sufficient precision the present-day kinematic parameters of the pulsar. The main difficulty here is related to the determination of the radial velocity, which is not known from observations. The process of integration of the trajectory is straightforward because the Galactic gravitational potential is known with sufficient precision. However, the problem arises because we do not know the pulsar's birthplace. Several assumptions can be made, and we discuss this topic below in detail. Mainly due to the uncertainty in the birthplace and the assumptions about radial velocities, the attribution of the pulsar to one of the velocity distribution modes is non-trivial. In many cases, it can be done only with some probability. 

Various research groups used similar approaches to obtain kick velocities in the case of radio pulsars \cite{2013MNRAS.430.2281N} and other objects, e.g., binaries with black holes \citep{2019MNRAS.489.3116A} and neutron stars  \citep{2023MNRAS.521.2504O}. However, our approach differs in detail, as discussed below.

 Below, in each subsection, we describe one by one the key ingredients of our approach.
 We begin with the sample selection procedure. Then, we describe how the radial velocities are chosen and how the trajectories of the pulsars are calculated. Afterward, we discuss how the possible birthplaces of the pulsars are identified. Finally, it is specified how a pulsar is attributed to one of the velocity distribution modes. 

\subsection{Sample of radio pulsars}
We selected radio pulsars from the Australia Telescope National Facility (ATNF) Pulsar Catalogue
v2.7.0 \citep{2005AJ....129.1993M}. 
This is done according to the following criteria:
\begin{itemize}
\item Either there is a robust parallax measurement of the pulsar ($\varpi \ge 3 \sigma_\varpi$, here $\varpi$ is the parallax and $\sigma_\varpi$ is the standard deviation of the uncertainty in the parallax measurement) or the dispersion measure distance obtained with the YMW16 model \citep{2017ApJ...835...29Y} is less than 10 kpc;
\item There is an available proper motion estimate;
\item The spin period, $P$, and its first derivative, $\dot{P}$ are known;
\item $\dot{P} > 5 \, \times \, 10^{-18}$, in order to exclude recycled pulsars;
\item There is no evidence that the pulsar is a component of a binary system or belongs to a globular cluster;
\item The source is not classified as an anomalous X-ray Pulsar, Soft Gamma-ray Repeater, etc. 
\end{itemize}
This selection results in a sample of 202 pulsars.
Our selection procedure is similar to that of \cite{2025ApJ...989L...8D}, who used 197 pulsars in their study. 

\subsection{Integration of spatial trajectories}
 
First, we have to specify the pulsar's present-day parameters, which serve as initial conditions for the trajectory integration. 

Of 202 pulsars in our sample, 75 have a robust parallax measurement with $\varpi \ge 3\sigma_\varpi$, while the remaining 127 pulsars do not.
To estimate the distances to pulsars with robust parallax measurements we employ a formalism proposed by \cite{2017A&A...608A..57V}. Given an observed parallax $\varpi$, the posterior distribution of $D_\varpi$ is:

\begin{equation}
\begin{split}
p(D_{\varpi} \mid \varpi_{\text{obs}}) \propto \exp \left( -\frac{([D_{\varpi} / \text{kpc}]^{-1} - [\varpi_{\text{obs}} / \text{mas}])^{2}}{2[\sigma_{\varpi} / \text{mas}]^{2}} \right) 
\\
\times 
D_{\varpi}^{2} \,R^{1.9} \, \exp \left( -\frac{|z\,(D_\varpi)|}{0.33 \,\text{kpc}} - \frac{R\,(D_\varpi)}{1.70 \,\text{kpc}} \right)
\end{split}
\label{eq:parallax_distances}
\end{equation}
where $R$ and $z$ are the Galactocentric radius and height of the pulsar and are calculated from $D_\varpi$ via Eq.~(\ref{eq:coords}).
Employing this formalism, for each pulsar with a robust parallax measurement, 100 values of $D_\varpi$ are sampled.
For other pulsars, 100 values of dispersion measure distance $D_\text{DM}$ are sampled from a Gaussian distribution with a standard deviation of $0.2 \times D_\text{DM}$~\citep{2024ApJ...970...90D}. 
Any negative values of both $D_\varpi$ and $D_\text{DM}$ are resampled again until positive values are yielded.

After obtaining 100 distance values $D$ ($D_\varpi$ or $D_\text{DM}$ respectively), for each pulsar we use their Galactic coordinates $l$ and $b$ to obtain 100  present-day positions in Galactocentric cylindrical coordinates $R$, $\phi$, and $z$:

\begin{equation}
    \begin{split}
        R &= \sqrt{R_{\odot}^2 + D^2\cos^2{b}-2R_{\odot}D\cos{b}\cos{l}} \\
        \phi &= \arccos{\left(\frac{R^2 + R_{\odot}^2 -D^2\cos^2{b}}{2RR_{\odot}}\right)}\\
        z &= D\sin{b} + z_{\odot},
    \end{split}
    \label{eq:coords}
\end{equation}
where $z_{\odot}=20.8\,\text{pc}$ \citep{2019MNRAS.482.1417B} and $R_{\odot}=8.122\,\text{kpc}$ \citep{2018A&A...615L..15G} are the Galactocentric height of the Sun and the projected distance from the Galactic center, respectively.

Subsequently, we use the combination of distances and proper motions to obtain 100 transverse velocity vectors $\vec{v_t}$ for each pulsar. Each transverse velocity vector $\vec{v_t}$ is a projection of the 3D velocity onto the sky corrected for the Galactic motion of the Sun.
However, the radial velocity component $v_r$ cannot be constrained by observations. 
Thus, it must be inferred from a prior assumption about the pulsar's 3D velocity vector. 
We follow a method similar to that of~\cite{2024MNRAS.527.1101G} and assume the isotropic velocity orientation relative to the pulsar's local standard of rest (LSR). 
The choice of LSR isotropy over an alternative assumption of Galactocentric isotropy was shown in \cite{2024A&A...689A.348D}.
We discuss the differences in Sect.~\ref{sect:discussion:GC}.

For each value of the distance $D$, 
the line of sight to the pulsar and its transverse velocity vector $\vec{v_t}$ define a singular plane.
If the 3D-velocity vector is assumed to be isotropically distributed, then $\theta$, the angle between the line of sight and the 3D velocity vector in the LSR reference frame, would be uniformly distributed over $(0, \pi)$.

For each pulsar, we calculate $10^5$ trajectories, and a very small number of unphysically large radial velocities which correspond to $\left|\cot\theta\right| \rightarrow \infty$ produce numerical issues. To avoid these problems, we resample the angle $\theta$ if $\theta > 0.99 \pi$ or $\theta < 0.01\pi$. In total, only 2\% of the line-of-sight---velocity angles are resampled.
Such a restriction is less severe than many other examples in the literature. E.g., \cite{2019MNRAS.489.3116A} in their study limited radial velocities to just the largest measured velocities for similar systems (black hole binaries). 

This procedure differs from that used by~\cite{2024MNRAS.527.1101G}, in that they sampled values of $\cos\theta$ from $\text{Uniform}(-1,\,1)$.
Nevertheless, in Sect.~\ref{sect:discussion} we show that using the angle sampling method from~\cite{2024MNRAS.527.1101G} produces results very similar to our model.

Then, for each distance value (and thus for each $\vec{v_t}$), a pulsar's line-of-sight velocity is obtained using:

\begin{equation}
    v_r=\left|\vec{v}_t-\vec{v}_{\text{LSR},\,t}\right|\hspace{.3mm}\cot\theta+v_{\text{LSR},\,r.}
    \label{eq:infer_vr}
\end{equation}

Here, $\vec{v}_{\text{LSR},\,t}$ and $v_{\text{LSR},\,r}$ are the transverse and line-of-sight components of LSR velocity, which are obtained using the Galactic potential from \cite{2017MNRAS.465...76M} with \texttt{GalPot}\footnote{https://github.com/PaulMcMillan-Astro/GalPot\label{fn:GalPot}}.
Note that the LSR isotropy assumption is not the only choice, e.g., \cite{2026arXiv260300390L}; we discuss this assumption in Sect.~\ref{sect:discussion}.

Having sampled 100 present-day positions and $10^3$ 3D velocities for each pulsar, we generate $10^5$ ($10^2 \times 10^3$) position-velocity combinations in total.
Subsequently, we reverse the velocity vectors and use the Galactic potential from \cite{2017MNRAS.465...76M} with \texttt{GalPot}\footref{fn:GalPot} to integrate the pulsar's orbits in the Milky Way. 
For each pulsar, the integration limit is chosen to be 1.5 times its spin-down age $\tau_c=P/2\dot P$, assuming it to be the upper estimate of its real age.
Kinematic data are collected at intervals with a constant time step of 0.1 Myr.

\subsection{Birthplaces}

Typically, it is impossible to determine the precise birthplace of a pulsar.
Rare exceptions are: pulsars in supernova remnants and pulsars for which a birth cluster (or a component from the disrupted binary) is identified.
Thus, in our approach, to determine the kick velocity of a pulsar, several presumed places of birth are chosen. 
Then, a kick velocity $v_\text{kick}$ is determined 
in each possible birthplace for each trajectory. 

First, we assume that pulsars are born close to the Galactic plane, and we consider the point of the trajectory that is closest to it as a possible birthplace of the pulsar. 
If the pulsar's trajectory crosses the Galactic plane several times over the integration time interval, then all the Galactic disk intersections are considered as possible birthplaces for this trajectory. 
The possible birthplaces for all trajectories are then grouped (first intersection, second intersection, etc.), where each group of birthplaces comprises up to $10^5$ different kick velocity values. A group of birthplaces is rejected 
if the number of trajectories crossing the Galactic plane for this group is less than $10^4$.

Subsequently, we account for trajectories that do not cross the Galactic plane by collecting kick velocities for each trajectory at the point where $t=-\tau_c$ and form a separate possible group of birthplaces, which comprises all $10^5$ different kick velocity values. 
It is worth noting that the spin-down age of the pulsar as an estimate of its true age has been criticized. 
E.g., this is the case for PSR J0953+0755 (B0950+08), for which its characteristic age is greater than its kinematic and cooling age by a factor of $\sim8$~\citep{2019MNRAS.482.3415I}.
However, our results do not change significantly under alternative assumptions about pulsar ages less than $\tau_c$. 
We discuss this in more detail in Sect.~\ref{sect:discussion}.

\begin{figure*}
    \centering
    \subfloat[PSR J0157+6212\label{fig:traj1}]{
    \includegraphics[width=0.45\textwidth]{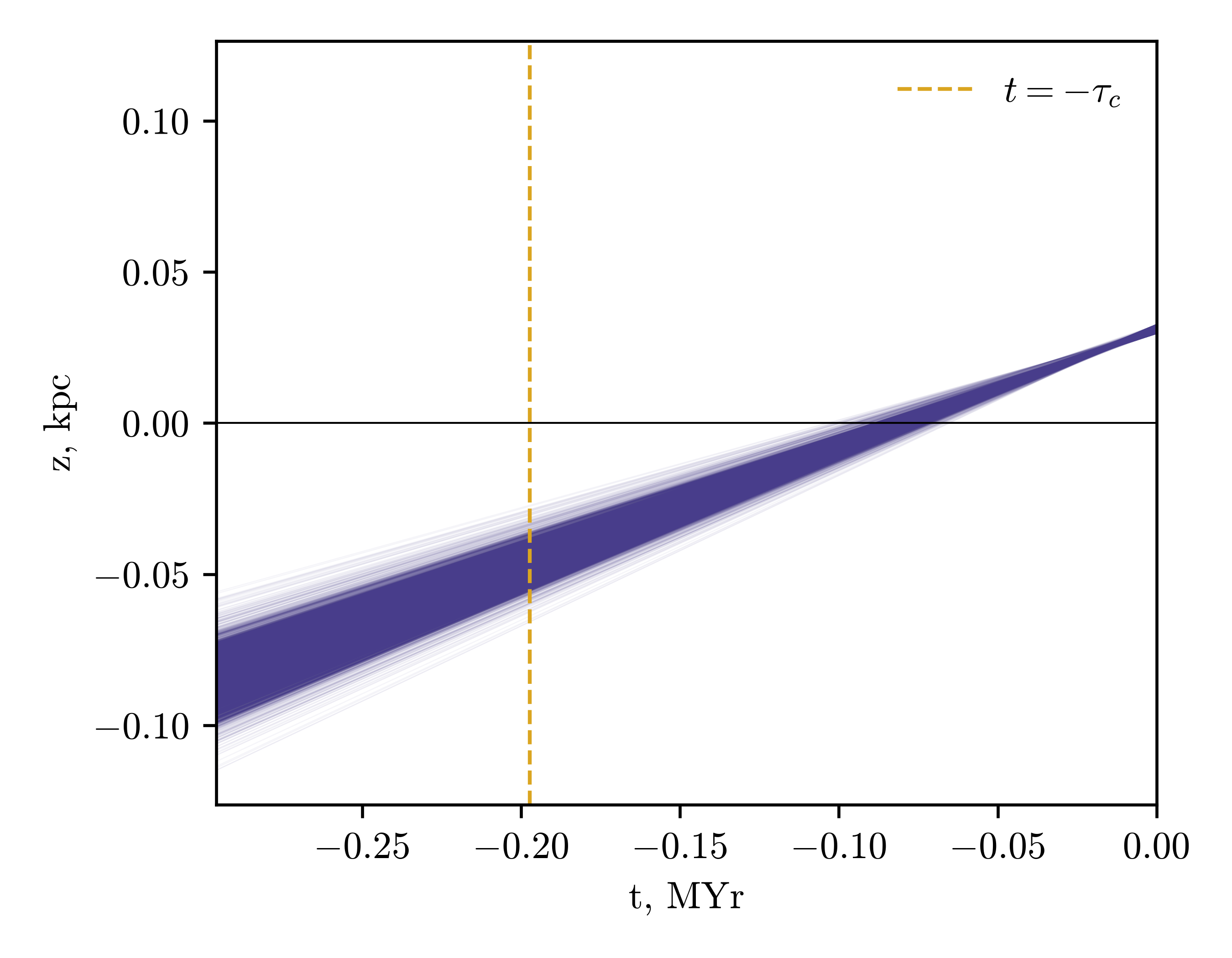}
    }
    \hfill
    \subfloat[PSR J0820-1350\label{fig:traj2}]{
        \includegraphics[width=0.45\textwidth]{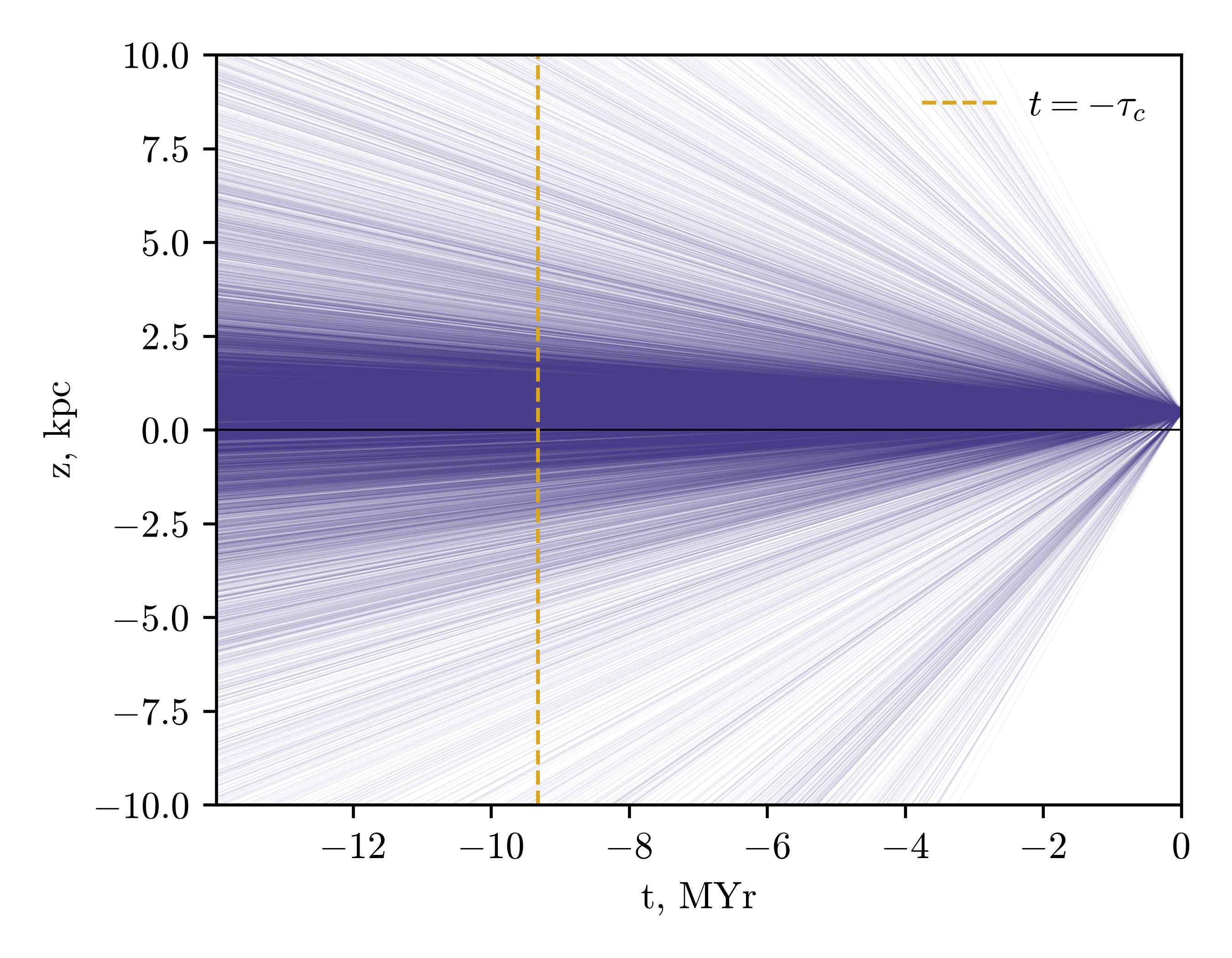}
    }
    \vspace{1em}
    \subfloat[PSR J2018+2839\label{fig:traj3}]{
        \includegraphics[width=0.45\textwidth]{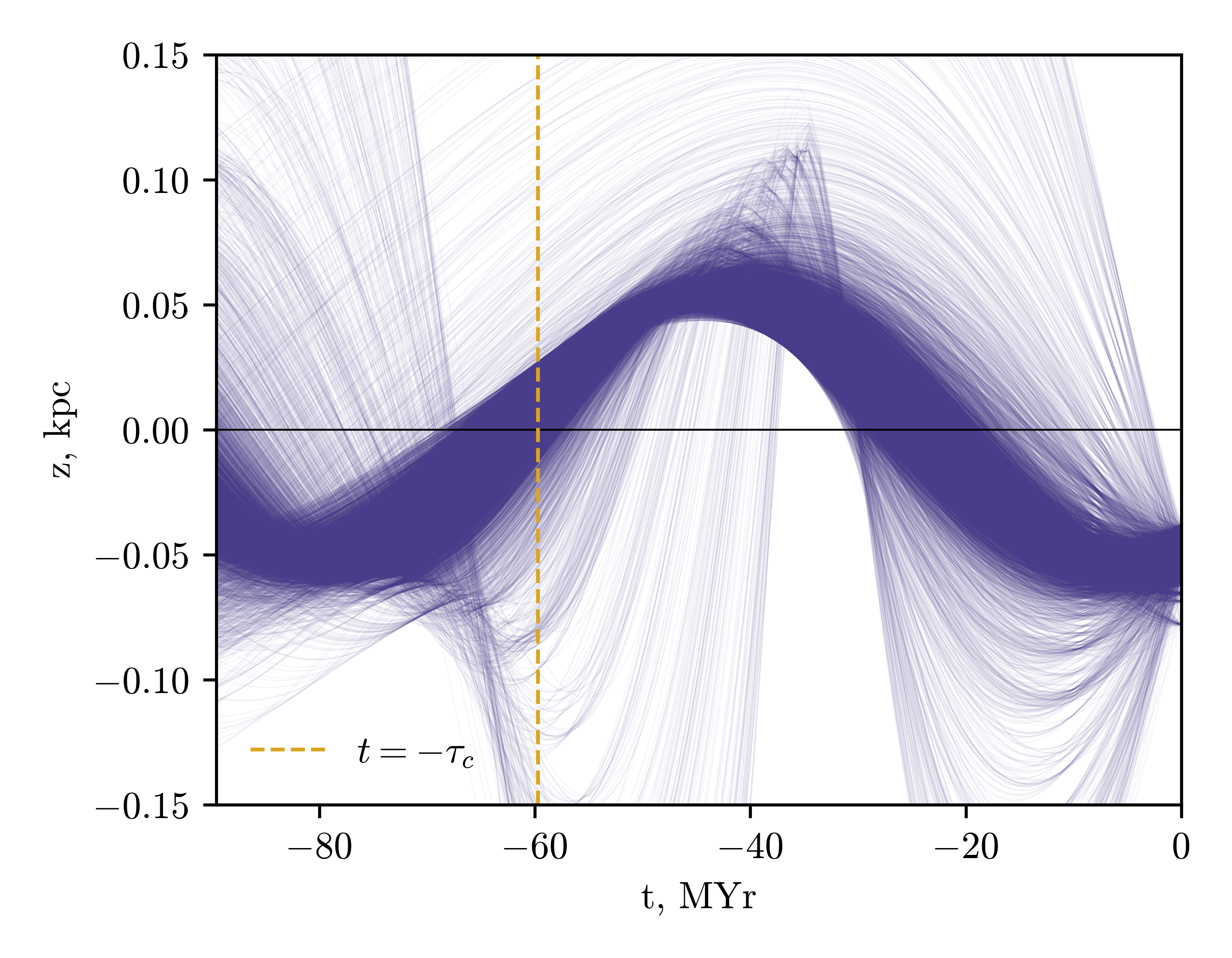}
    }
    \hfill
    \subfloat[PSR J1841-0425\label{fig:traj4}]{
        \includegraphics[width=0.45\textwidth]{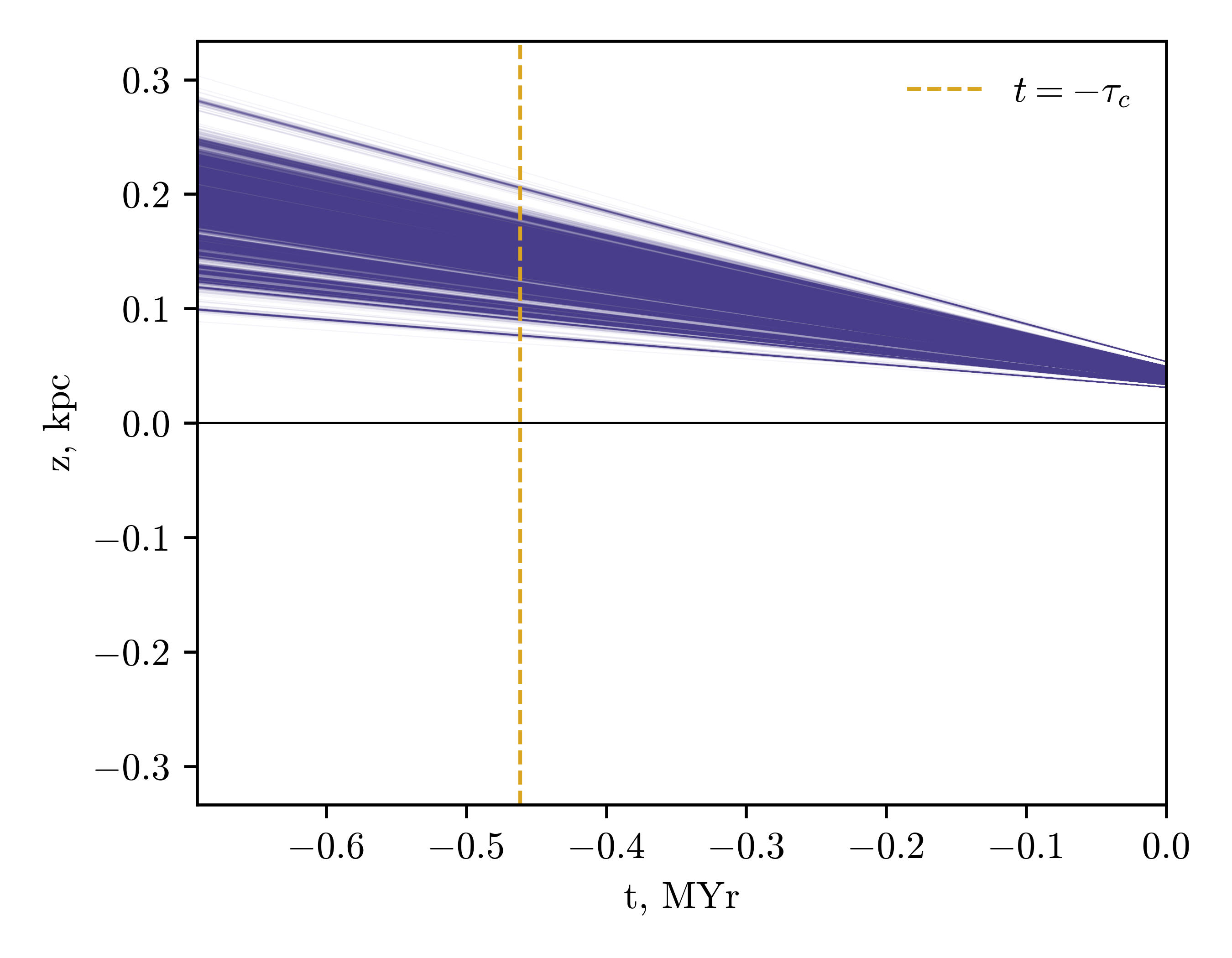}
    }
    \caption{Four typical scenarios representing pulsar trajectories. The time evolution of the $z$-coordinate is shown.
    In each panel, the horizontal axis is terminated on the left at $t=-1.5\tau_c$.
    Each trajectory terminates at its present-day position at the right edge of the horizontal axis.
    The vertical dashed yellow line represents $t=-\tau_c$. For clarity purposes, only every tenth trajectory is shown at each panel.}
    \label{fig:trajectories}
\end{figure*}

For each pulsar, the set of trajectories falls within one of several scenarios, shown in Figure~\ref{fig:trajectories}.
The most common scenario among our pulsar sample (especially for large velocities) is the one depicted in Panel~\ref{fig:traj1}: all of the trajectories cross the Galactic plane only once, which, combined with the birthplace corresponding to $t=-\tau_c$, results in 
two possible birthplaces for each trajectory. 
In this case, each of the two birthplaces encompasses all $10^5$ trajectories. 
Alternatively, some of the trajectories might not cross the Galactic plane within the integration interval, as shown in Panel~\ref{fig:traj2}. 
In this scenario, the pulsar also has two possible birthplaces per trajectory. However, the birthplace associated with the Galactic plane encompasses only a fraction of all $10^5$ trajectories.
Panel~\ref{fig:traj3} shows a scenario with very complicated behavior when some trajectories cross the Galactic plane at various places several times, while others do not cross it at all.
Lastly, Panel~\ref{fig:traj4} shows a scenario in which none of the trajectories cross the Galactic plane, which therefore results in the pulsar having only one assumed birthplace (corresponding to $t=-\tau_c$), encompassing all $10^5$ trajectories.

Finally, under the assumption that the kick velocity is isotropic in the local standard of rest at birth, we evaluate the kick velocity for each trajectory and birthplace as: 
\mbox{$v_\text{kick} = \left| \vec{v} - \vec{v}_\text{LSR}\right|$}, where $\vec{v}$ and $\vec{v}_\text{LSR}$ are the pulsar's 3D velocity and the LSR velocity at the birthplace, respectively.

\subsection{Pulsar categorization}\label{sect:model:pulsar_categorization}
Our goal is to separate our pulsar sample into two distinct groups, representing low-velocity and high-velocity modes of a prior kick velocity distribution, in order to reveal the differences in the pulsar parameters of the two groups. 
As shown by~\cite{2020MNRAS.494.3663I}, this kick velocity distribution is well described by a weighted sum of two Maxwellian distributions. 
We, therefore, use two Maxwellian distribution functions to represent the natal kick components corresponding to the low- and high-velocity groups. 
In~\cite{2020MNRAS.494.3663I}, the authors provide the kick velocity distribution based on studies of young pulsars (spin-down age less than 3 Myr). 
We use these parameters in our study:
$w_1=0.2$, the fraction of low-velocity pulsars, $\sigma_1=56$ \kms  
for the low-velocity component, and $\sigma_2=336$~\kms
for the high-velocity component.
Notably, \cite{2021MNRAS.508.3345I} revisited the kick distribution by~\cite{2020MNRAS.494.3663I} by incorporating binary systems with observations of Galactic Be X-ray binaries, and updated the parameter of the low-velocity mode to be $\sigma_1 = 45$~\kms.
We, however, use the original distribution by~\cite{2020MNRAS.494.3663I} 
based on the analysis of the sample of isolated radio pulsars, as we also use only this type of source.

Having obtained the kick velocities for each possible pulsar birthplace, we compute the probability of belonging to either the low-velocity or the high-velocity mode of the chosen prior kick distribution. 
For each pulsar, this probability is proportional to the sum of individual probabilities for each $v_\text{kick}$ calculated using the Maxwellian distribution, corresponding to the two modes of kick distribution. 
\begin{equation}
    p_{1,2} \propto w_{1,2}\sum_{i=1}^{N_\mathrm{v}}{ f_{1,2}(v_i)}
    \hspace{1cm} 
    p_1+p_2 = 1
\label{eq:coefficients}
\end{equation}
Here, $f_\text{1,2}$ are Maxwellian distribution modes of the prior kick velocity distribution with corresponding parameters $\sigma_1=56$~\kms, $\sigma_2=336$~\kms, 
$w_1=0.2$, and $w_2 = 1-w_1=0.8$ are weights of the two distribution modes, 
and $N_\mathrm{v}$ is the number of the kick velocity values for a particular birthplace.
If a particular kick velocity value is lower than a critical value $v_\text{crit} \simeq 160$~\kms (where $w_1f_1(v_\text{crit}) = w_2f_2(v_\text{crit})$), 
then its contribution to the low-velocity mode probability,  $p_1$, is greater than to the high-velocity mode probability, $p_2$, and vice versa.
This method is similar to the evaluation of a likelihood function, which is often a product of probability densities. 
Note that a kick velocity distribution obtained for a particular birthplace differs in shape significantly from both of the Maxwellian distributions.
In this case, the product of probability densities is extremely close to zero even if the peak of the obtained kick velocity distribution is close to the peak of one of the Maxwellian distributions. 

The described procedure ensures that all pulsars have an estimated probability of belonging to either distribution mode at each of their possible birthplaces.
If a given pulsar belongs to either the low- or high-velocity mode of the distribution with a probability higher than 0.5 at each of its possible birthplaces, we categorize the pulsar to belong to the low-velocity mode or the high-velocity mode, and we consider the distribution mode to be "unambiguously determined".
Otherwise, the pulsar is categorized as belonging to the distribution mode, the probability of which is greater than 0.5 at the majority of the pulsars' birthplaces, and the chosen distribution mode is considered to be determined "ambiguously".
Analysis only for the unambiguously categorized pulsars is presented alongside the full sample (i.e. both ambiguously and unambiguously categorized pulsars) in Table.~\ref{tab:KS}. 
Figures~2--7 show the full pulsar sample, not taking into account whether a pulsar is categorized ambiguously or unambiguously.


\section{Results}\label{sect:results}

In this section, we describe the results obtained from our analysis. 

\begin{figure*}
    \centering
    \subfloat[All pulsars in our sample\label{fig:kicks_all}]{
    \includegraphics[width=0.45\textwidth]{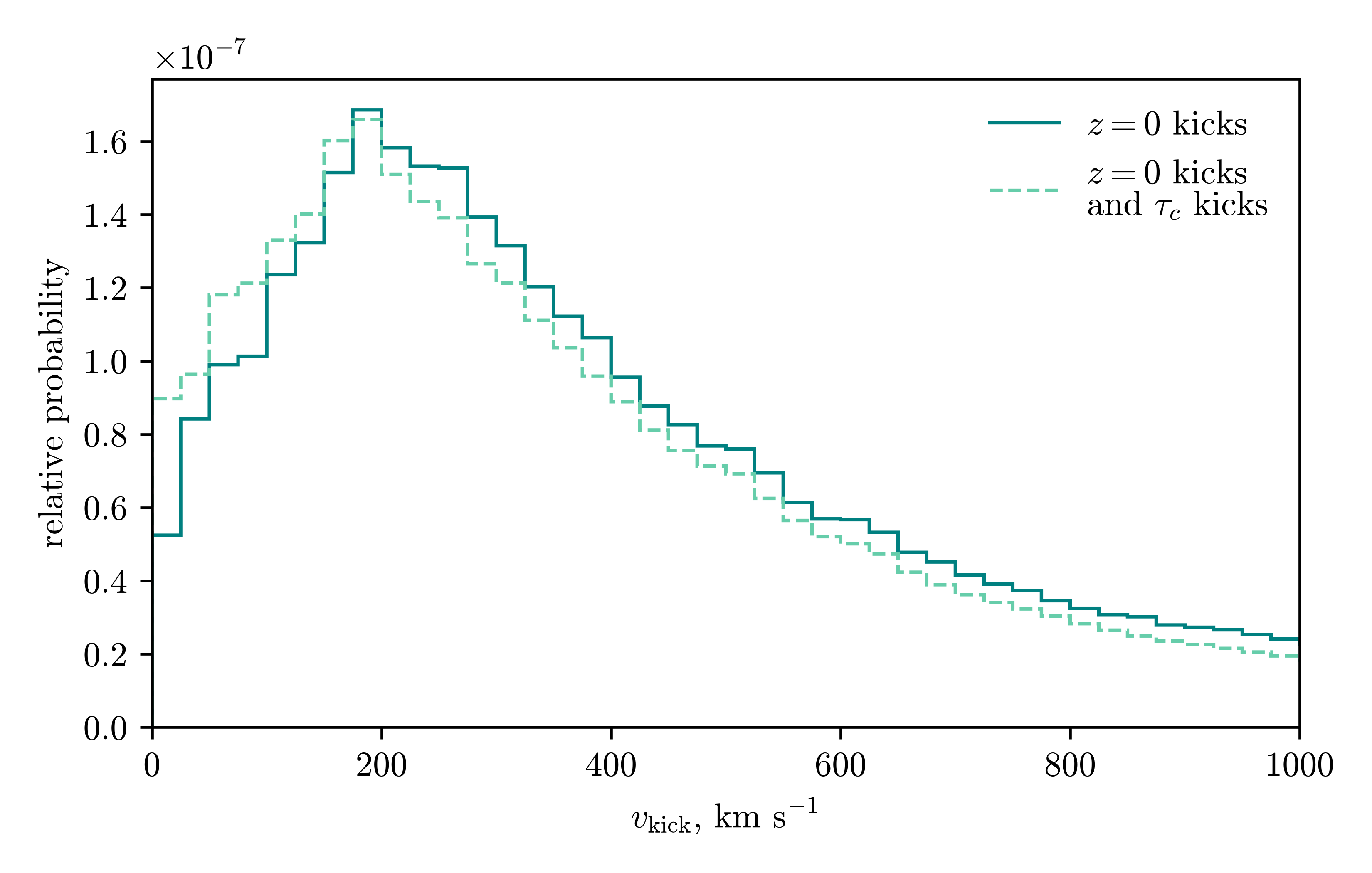}
    }
    \hfill
    \subfloat[Pulsars with $\tau_c \le 10$~ Myr\label{fig:kicks_young}]{
    \includegraphics[width=0.45\textwidth]{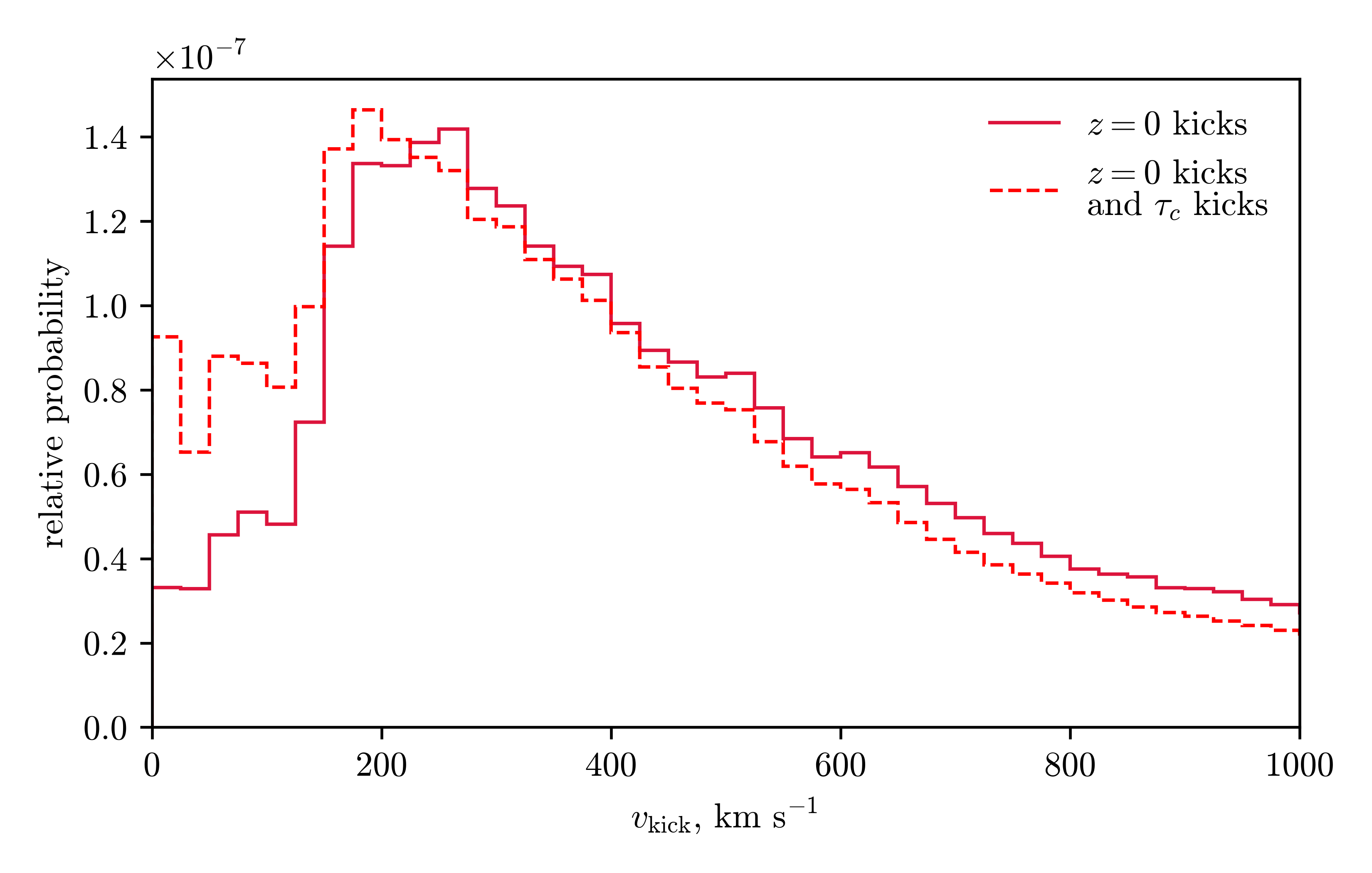}
    }
    \caption{Kick velocity distribution for our sample of radio pulsars. In Panel~\ref{fig:kicks_all} the distribution is shown for all pulsars. While in Panel~\ref{fig:kicks_young} only young pulsars with characteristic ages $\tau_c \le 10$~Myr are used. In both panels, kick velocities obtained at points where a given trajectory crosses the Galactic plane ($z=0$) are shown as a step histogram with a solid darker line, whereas kicks obtained either under the condition $z=0$ or $t=-\tau_c$ are shown with a dashed step histogram.}
    \label{fig:kick_distributions}
\end{figure*}

First, as an illustration, we present the distribution of natal kicks obtained using the method described above. In Fig.~\ref{fig:kick_distributions}, we plot kick velocities for the whole sample (Panel~\ref{fig:kicks_all}) and for young pulsars (Panel~\ref{fig:kicks_young}). 
For each pulsar we do not define a single kick velocity value, but use all (up to $10^5$) trajectories and all possible birthplaces producing a normalized probability density distribution. 
Then, all individual pulsar distributions are summed and renormalized to obtain the population kick velocity distribution. 
In each panel, we show two curves as birthplaces are uncertain.
The curves, shown as a lighter dashed line, represent the results obtained under the assumption that all pulsars are born at the Galactic plane.
These curves are obtained by using only kicks at points where a given trajectory crosses the Galactic plane ($z=0$). 
Pulsars for which none of the trajectories cross the Galactic plane are ignored.
Darker solid curves are plotted assuming that, for some pulsars, the true age is equal to the characteristic one. 
For these lines, kicks from both $z=0$ birthplaces and $t=-\tau_c$ are used.
However, our goal is not to derive the exact shape of the kick velocity distribution.
Below, under the assumption of bimodality of the distribution, we separate pulsars into two groups and analyze their properties in each mode. 

\subsection{Low- and high-velocity subsamples of radio pulsars}

As a result of our analysis, the sample of 202 pulsars is separated into two distribution modes:

\begin{itemize}
    \item 46 pulsars belong to the low-velocity mode, of which 40 are categorized unambiguously, and 6 are categorized ambiguously;
    \item 156 pulsars belong to the high-velocity mode, of which 155 are categorized unambiguously, and only one is categorized ambiguously.
\end{itemize}

Thus, $\sim23$ percent of pulsars belong to the low-velocity mode. 
This figure is well
within a $1\sigma$ confidence interval of
that obtained by \cite{2020MNRAS.494.3663I} for their young pulsar sample: $20^{+11}_{-10}$ per cent.


In the following, we first show the differences in some parameters between the two modes, which arise naturally from selection bias in the pulsar observations. 
Then, we show that for some quantities, which are not subject to strong selection, there is no clear difference between the two modes.
Finally, we show differences between the two modes that most probably cannot be attributed to selection bias alone. 

Results of the Kolmogorov-Smirnov test for these parameters are shown in Table~\ref{tab:KS}.
We discuss in detail and propose possible explanations for such discrepancies in Sect.~\ref{sect:discussion}.

\begin{figure}[h]
\centering
\includegraphics[width=0.45\textwidth]{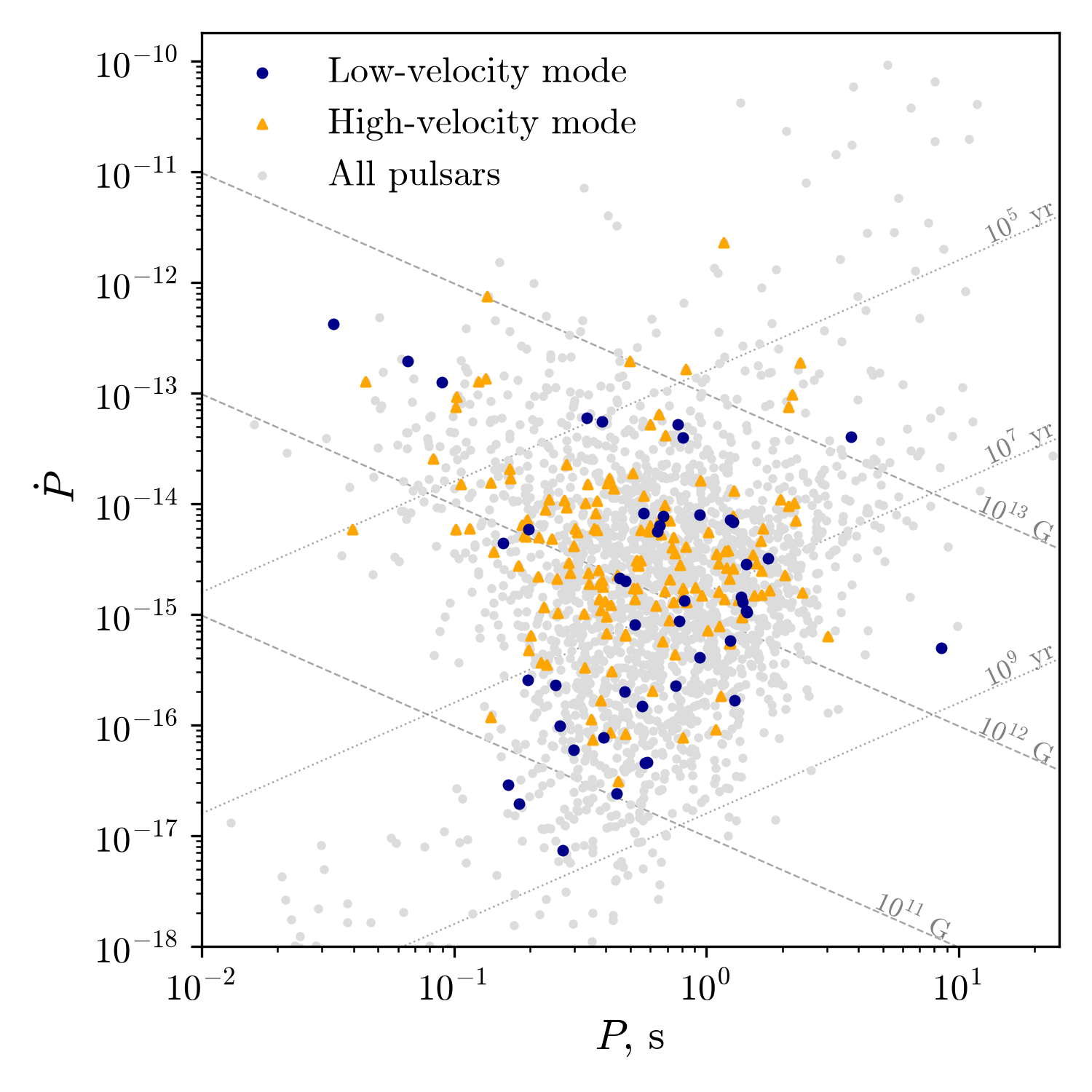}
\caption{The $P-\dot{P}$ diagram. Pulsars from the low- and high-velocity distribution modes are represented by dark blue and orange circles, respectively. 
Grey points represent all sources from the ATNF Pulsar Catalogue with measured $P$ and $\dot{P}$.
}\label{fig:PPdot_mixed}
\end{figure}

\begin{figure*}
    \centering
    \subfloat[Spin-down age\label{fig:AGE_CDF}]{
    \includegraphics[width=0.45\textwidth]{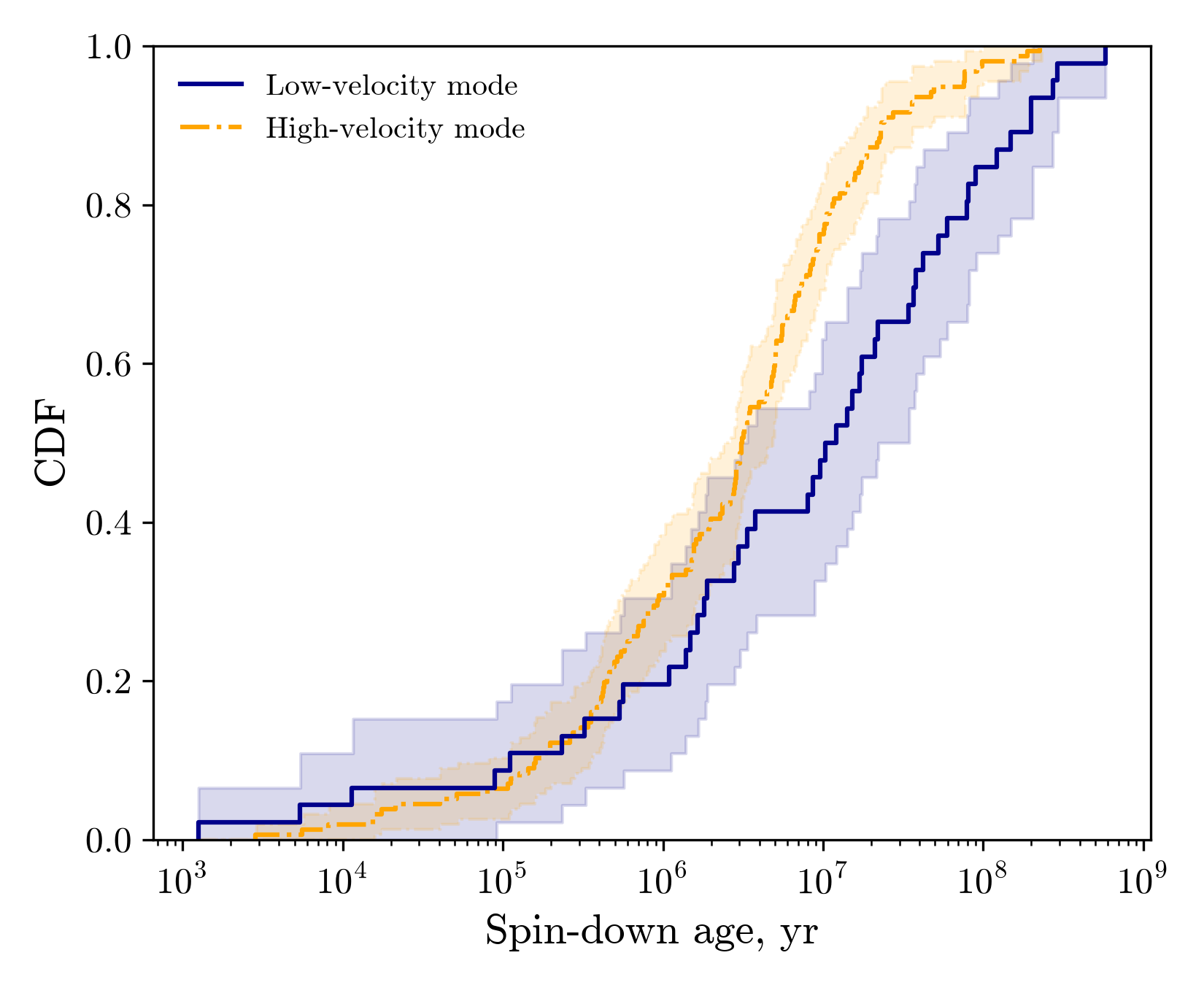}
    }
    \hfill
    \subfloat[Heliocentric distance\label{fig:fig:DIST_CDF}]{
    \includegraphics[width=0.45\textwidth]{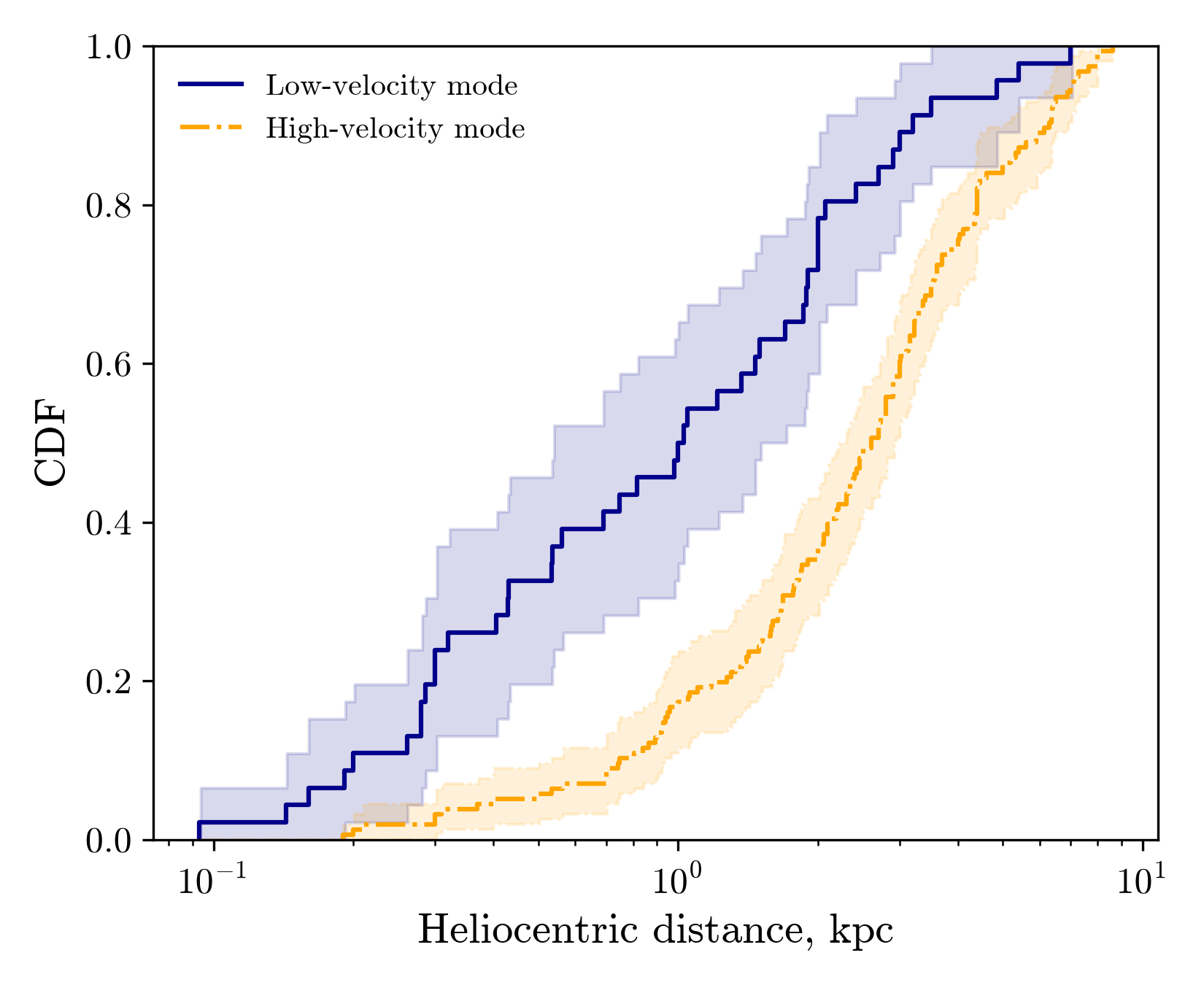}
    }
    \caption{Cumulative distribution functions of pulsar spin-down age and heliocentric distance. Low-velocity and high-velocity distribution modes are represented by dark blue and orange lines, respectively. Shaded areas represent a bootstrapped 95\% confidence interval. The horizontal axes are given in a logarithmic scale. The CDFs are shown for the full pulsar sample.}
    \label{fig:CDFs}
\end{figure*}

Pulsars, categorized into the two distribution modes, are shown in Figure~\ref{fig:PPdot_mixed}. Already, the $P-\dot{P}$ diagram clearly demonstrates a selection bias: pulsars in the high-velocity mode, on average, have a higher rotational energy-loss rate and therefore appear higher on the diagram. This is related to the fact that high-velocity pulsars are on average more distant, see Fig.~\ref{fig:CDFs}.

\begin{table}[h]
\centering
\caption{
Kolmogorov-Smirnov $D$-statistic and corresponding $p$-values for parameters of low-velocity and high-velocity pulsars.
The results for both the full sample and only the unambiguously categorized pulsars are shown.
}
\label{tab:KS}
\setlength{\tabcolsep}{4pt}
\begin{tabular}{@{}lllll@{}}
\toprule
          & \multicolumn{2}{c}{Full sample} & \multicolumn{2}{c}{\makecell[l]{Unambiguously\\categorized}} \\ \cmidrule(lr){2-3} \cmidrule(l){4-5}
Parameter & $D$-statistic    & $p$-value       & $D$-statistic      & $p$-value       \\ \midrule
Spin-down age & $0.308$ & $1.73\times10^{-3}$ & $0.318$ & $2.37\times10^{-3}$ \\
\makecell[l]{Heliocentric\\distance} & $0.420$ & $3.74\times10^{-6}$ & $0.419$ & $1.48\times10^{-5}$ \\
W10 & $0.129$ & $0.571$ & $0.114$ & $0.765$ \\
W50 & $0.143$ & $0.416$ & $0.118$ & $0.719$ \\
$B_\text{surface}$ & $0.217$ & $0.058$ & $0.229$ & $0.059$ \\
Radio luminosity & $0.403$ & $1.72\times10^{-5}$ & $0.409$ & $4.15\times10^{-5}$ \\
\bottomrule
\end{tabular}
\end{table}

For the same reason, pulsars from the two modes demonstrate different distributions in the spin-down age, see
Figure~\ref{fig:AGE_CDF}.
High-velocity pulsars are younger, on average, than the low-velocity pulsars.
All these differences between the two modes arise from the same selection bias in pulsar observations: pulsars with well-measured low velocities are situated at smaller distances.
There are two reasons for this bias. At first, it is more difficult to measure small velocities at large distances. In addition, pulsars with large kicks leave the local volume relatively quickly. 
This bias is well-known. Thus, it was expected to obtain the characteristics shown in Figs.~\ref{fig:PPdot_mixed} and \ref{fig:CDFs}. 
Notably, the distance-dependent kinematic bias can be overcome by using a prior relationship between natal kicks and observed distances. This was explicitly shown by \cite{2026ApJ..1000L..56D}.
Results of the Kolmogorov-Smirnov test for spin-down age $\tau_c$ and heliocentric distance $D$ are presented in Table~\ref{tab:KS}. The difference in these parameters between the two modes is clearly demonstrated by the $p$-values  $1.73\times10^{-3}$ and $3.74\times10^{-6}$ for the full sample, and $2.73\times10^{-3}$ and $1.48\times10^{-5}$ for unambiguously categorized pulsars, respectively.

\begin{figure*}
    \centering
    \subfloat[Pulse width at 10\% maximum\label{fig:W10_CDF}]{
    \includegraphics[width=0.45\textwidth]{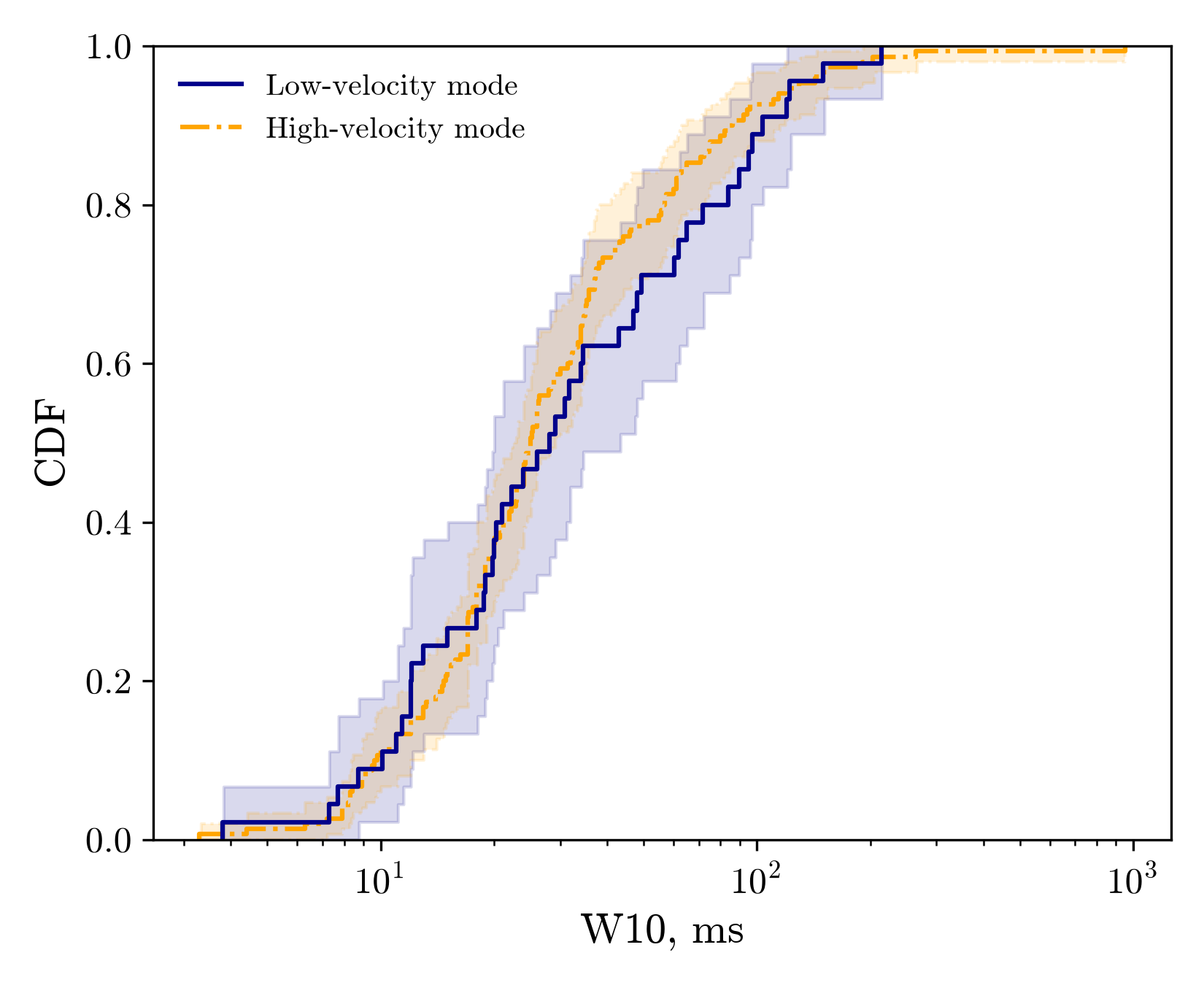}
    }
    \hfill
    \subfloat[Pulse width at 50\% maximum\label{fig:W50_CDF}]{
    \includegraphics[width=0.45\textwidth]{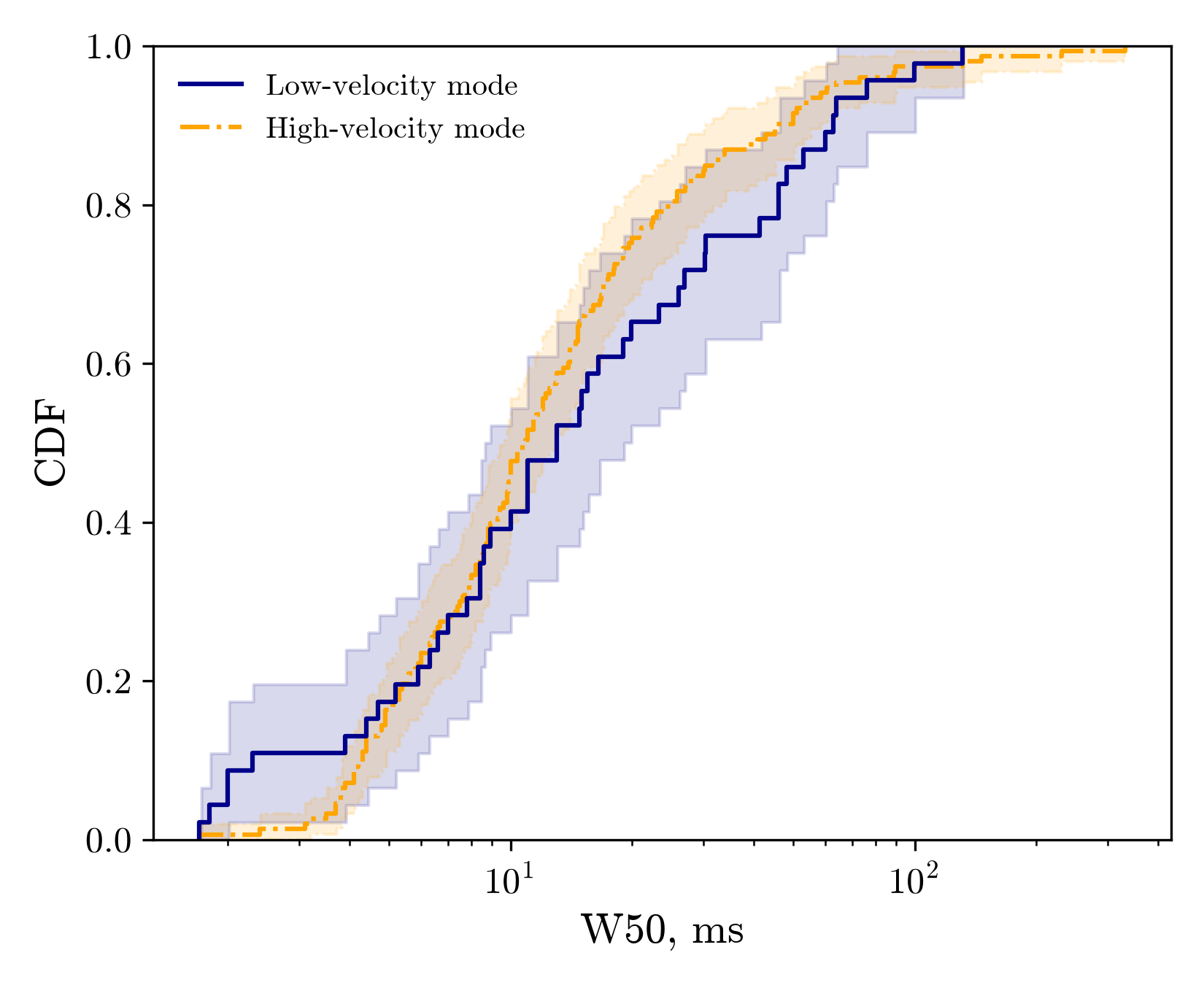}
    }
    \caption{Cumulative distribution functions of pulse width at 10\% and 50\% of the maximum intensity. Low-velocity and high-velocity distribution modes are represented by dark blue and orange, respectively. 
    Shaded areas represent a bootstrapped 95\% confidence interval. 
    The horizontal axes are on logarithmic scales. The CDFs are shown for the full pulsar sample.}
    \label{fig:W_CDFs}
\end{figure*}

Figures~\ref{fig:W10_CDF}~and~\ref{fig:W50_CDF} show cumulative distributions of pulse width W10 and W50, respectively. 
For W10 and W50, Kolmogorov-Smirnov test fails rejecting the null hypothesis at $p\gtrsim0.4$ for both the full sample, and the unambiguously categorized pulsar, see Table~\ref{tab:KS}.

Pulse width is not subject to strong selection and, therefore, pulsars from low- and high-velocity modes have similar pulse width distributions.

\subsection{Magnetic fields of pulsars in the two modes of the kick velocity distribution}

\begin{figure}[h]
\centering
\includegraphics[width=0.4\textwidth]{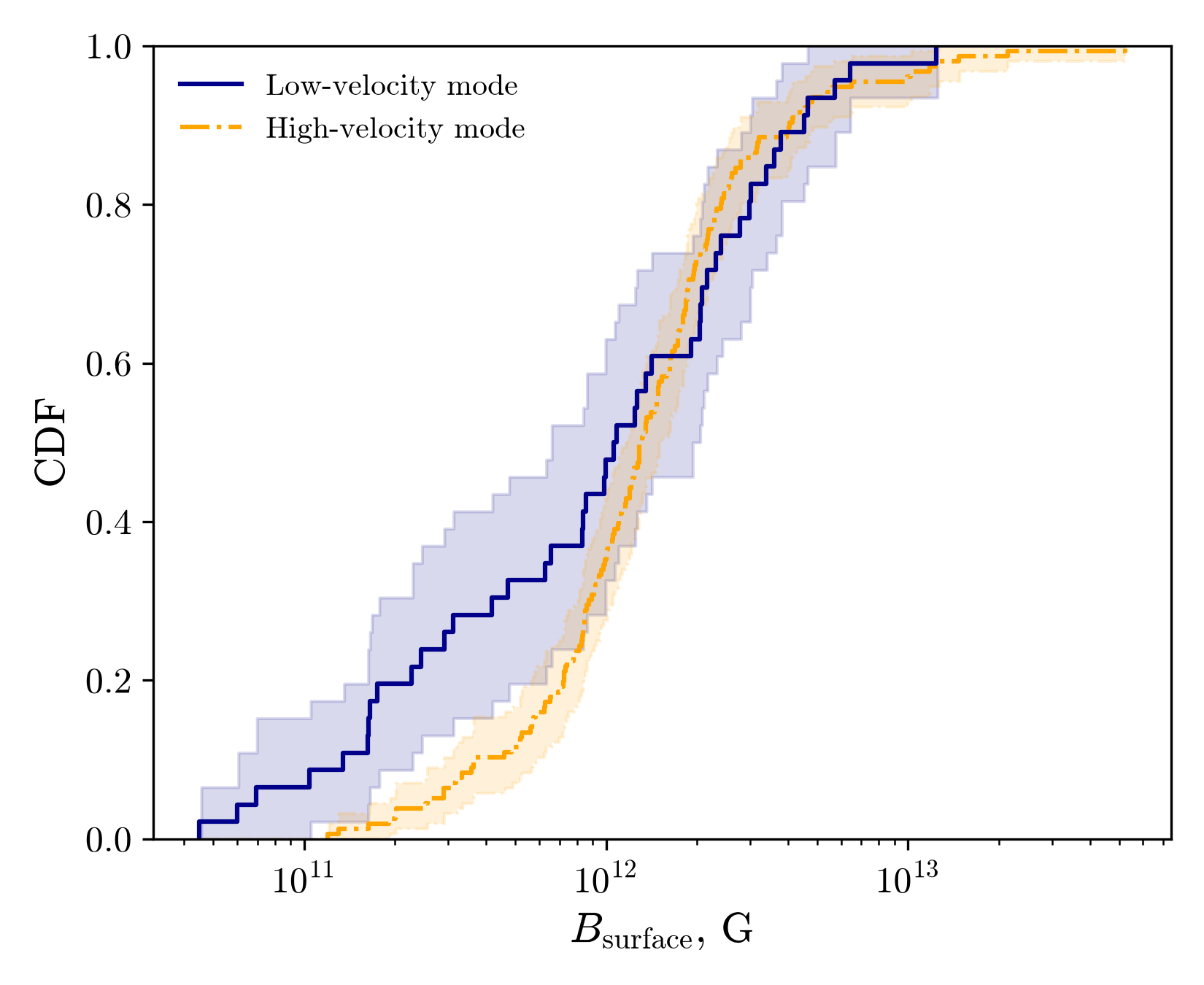}
\caption{Cumulative distribution functions of the effective surface dipolar magnetic field.
Low-velocity and high-velocity distribution modes are represented by dark blue and orange, respectively. 
Shaded areas represent a bootstrapped 95\% confidence interval. 
The horizontal axis is on a logarithmic scale. The CDFs are shown for the full pulsar sample.
}\label{fig:BSURF_CDF}
\end{figure}

Figure~\ref{fig:BSURF_CDF} shows the difference in the distributions of the effective surface magnetic fields of NSs in the two kick velocity modes. 
In this plot, we used the magnetic field values taken from the ATNF online catalogue.
Pulsars with lower magnetic fields ($\lesssim 10^{12}$~G) are more common in the low-velocity mode. 
The distributions converge for pulsars with magnetic fields larger than $10^{12}$~G.

The Kolmogorov-Smirnov test yields a $p$-value of $\simeq0.06$ for both the full pulsar sample and the subsample of unambiguously categorized pulsars, indicating barely a statistically significant difference between the magnetic-field distributions of the two modes at the $\alpha=0.1$ significance level.

\begin{figure}[h!]
\centering
\includegraphics[width=0.45\textwidth]{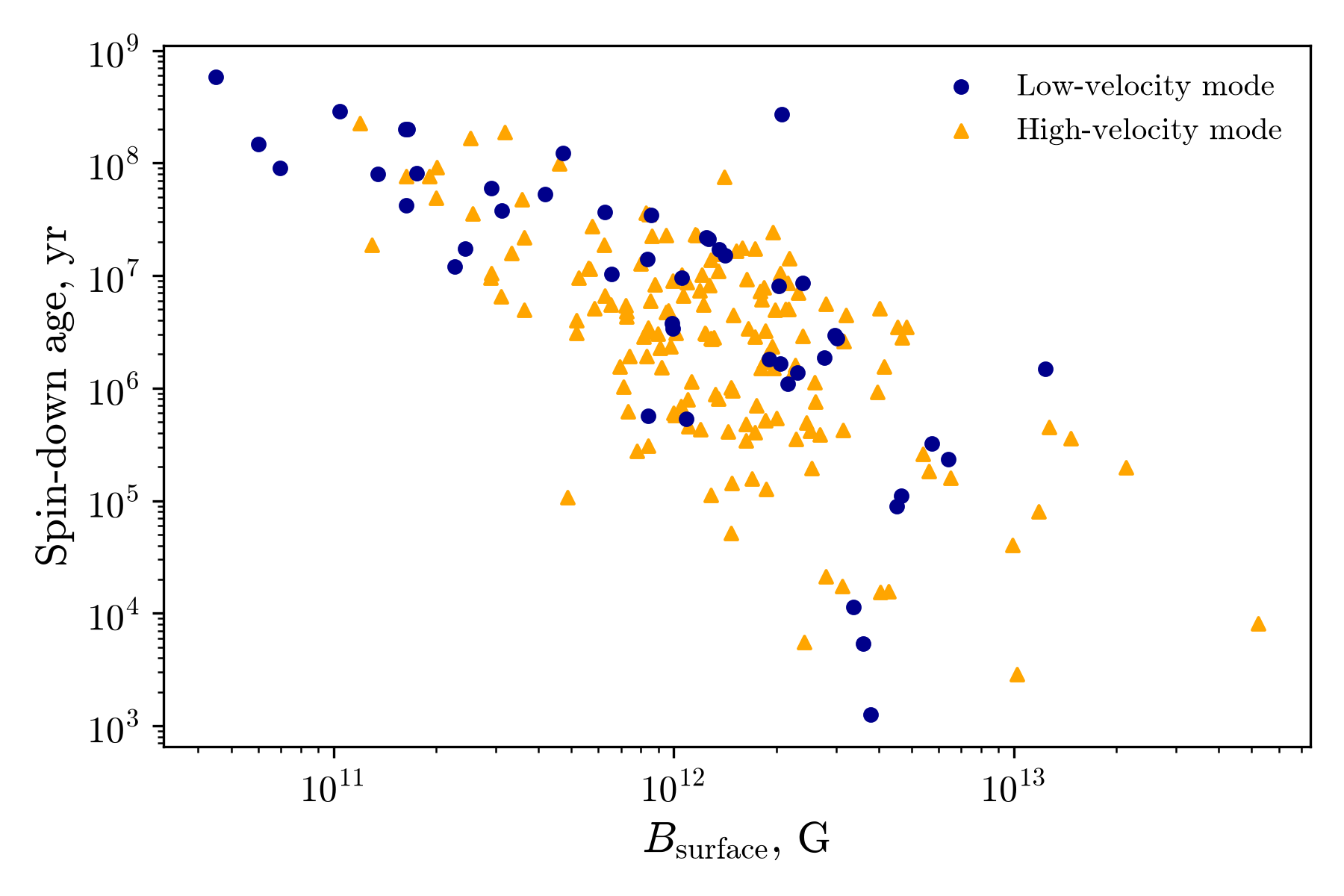}
\caption{Effective surface dipolar magnetic field plotted against the spin-down age. 
Low-velocity and high-velocity distribution modes are represented by dark blue circles and orange triangles, respectively.
}\label{fig:B_surf_t}
\end{figure}

This discrepancy can be partially attributed to selection bias, as younger pulsars have stronger magnetic fields 
(see Fig.~\ref{fig:B_surf_t}) and pulsars with higher fields, on average, have higher luminosities. 
Yet, it can also be explained by the difference in the physical pulsar parameters of the two distribution modes, see Sect.~\ref{sect:discussion}.


For illustrative purposes, we provide a scatter plot of the magnetic fields against their Galactocentric distances.
Pulsars with smaller distances mostly belong to the low-velocity mode, whereas pulsars further away mostly belong to the high-velocity mode.
The difference in distances is explained by selection bias.
Identically to Fig.~\ref{fig:BSURF_CDF}, most of the pulsars with magnetic fields $\lesssim 10^{12}$~G belong to the low-velocity mode. 
Moreover, all pulsars with magnetic fields $\lesssim  10^{11}$~G are categorized into the low-velocity mode. However, it is necessary to increase the statistics of low-field pulsars to obtain robust constraints on their distribution.

\subsection{Braking indices}

\begin{figure}[h]
\centering
\includegraphics[width=0.45\textwidth]{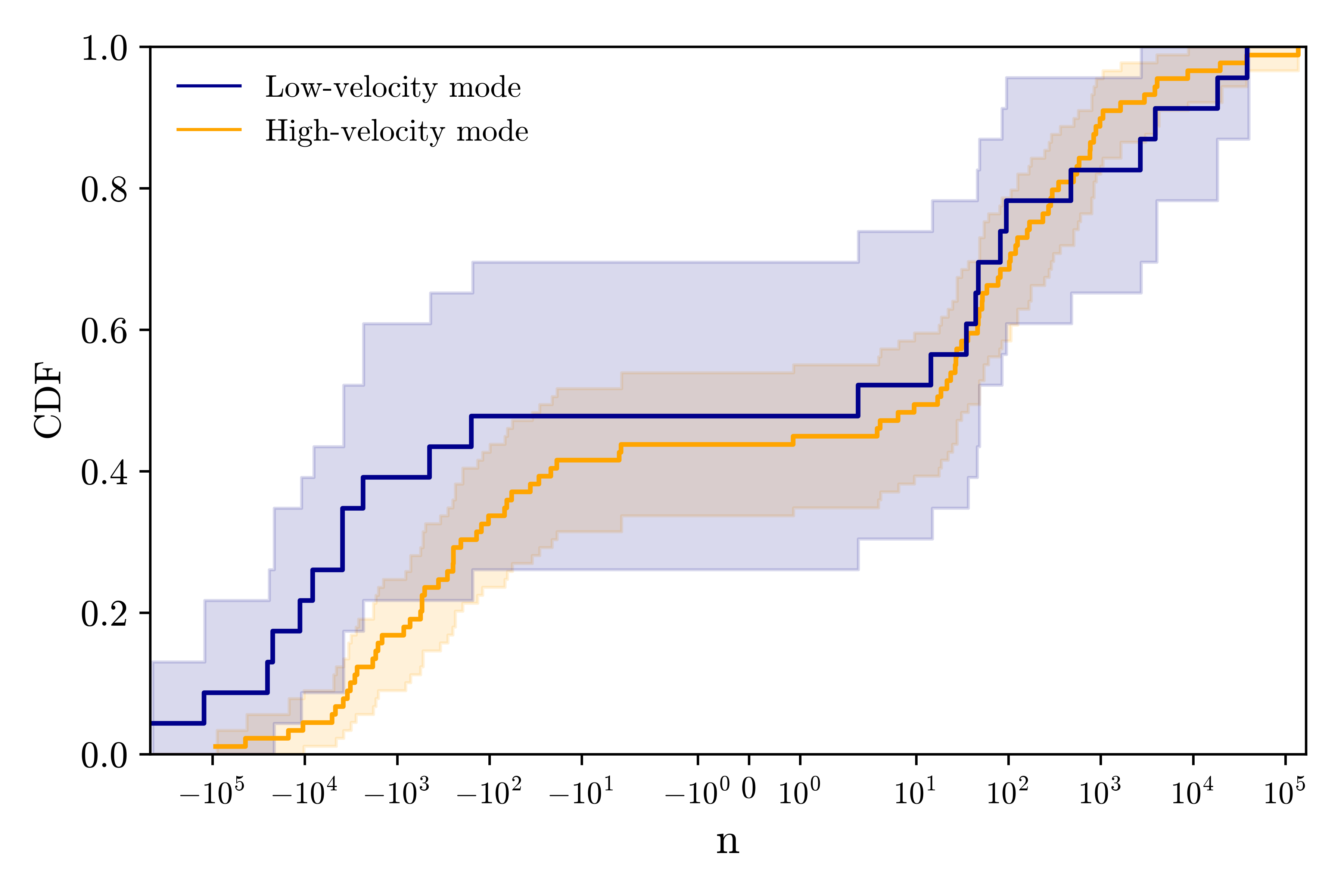}
\caption{
Cumulative distribution functions of braking indices of pulsars. 
Low-velocity and high-velocity distribution modes are represented by dark blue and orange, respectively.
Shaded areas represent a bootstrapped 95\% confidence interval. 
The horizontal axis is on a symmetric logarithmic scale.
}\label{fig:n}
\end{figure}

The difference between the distribution of braking indices of pulsars from the low- and high- velocity modes is shown in Figure~\ref{fig:n}. 
Braking indices are defined as $n = (\nu\ddot{\nu})/\dot{\nu}^2$, where $\nu$, $\dot\nu$, and $\ddot\nu$ are the pulsar's frequency and its first and second derivatives. 
We evaluate braking indices using values for $\nu$, $\dot\nu$, and $\ddot\nu$ from the ATNF Pulsar Catalogue. 
Only pulsars with $\left|\ddot{\nu}\right| > 3\sigma_{\ddot{\nu}}$ are shown in Figure~\ref{fig:n}, where $\sigma_{\ddot{\nu}}$ is the error of the second derivative of the frequency taken from the ATNF Pulsar Catalogue. 
In total, 112 pulsars (89 high-velocity pulsars and 23 low-velocity pulsars, of which 2 are categorized ambiguously) with well-measured second frequency derivatives are shown.
The two cumulative distributions are very similar. 
The only noticeable difference is that the fraction of low-velocity pulsars with negative braking indices is larger in comparison with high-velocity pulsars.
However, this can be due to low statistics of low-velocity pulsars with $n<0$.

\subsection{Spin-velocity alignment}

In our modeling, we ignore any effects resulting from the spin-velocity alignment. 
This correlation was initially proposed for the Crab \citep{1999A&A...344..367C} and Vela \citep{2001ApJ...556..380H} pulsars. 
Later on, it was demonstrated for a large sample of objects, see \citep{2005MNRAS.364.1397J, 2007MNRAS.381.1625J, 2007ApJ...664..443R, 2012MNRAS.423.2736N, 2015ApJ...804..112R}. Using polarization data, it was shown that the spin axis and the spatial velocity vector are not directed randomly relative to each other, but there is a tendency for them to co-align. Alignment was also demonstrated for NSs in binary systems  \citep{Valli:2025}.

In \cite{2003ApJ...585L..41R, 2004ApJ...601..479N}, the spin-velocity alignment was demonstrated in several cases using 3D data. Finally, the most robust proof of the alignment was shown by \citep{2021NatAs...5..788Y} for PSR J0538+2817.  
Recently, \cite{2023ApJ...944..153M} suggested that this property results in a selection effect for radio pulsars with small and large magnetic obliquities. 
\cite{2025PASA...42..106B} studied this effect in more detail. 
Theoretical interpretations for the origin of this correlation in frameworks of different supernova mechanisms were proposed by several authors; see, e.g., \cite{2026arXiv260617496L} and references therein.
They demonstrated that peculiar velocities of weakly and strongly oblique pulsars are distributed differently. 

We assume that the effect is not strong enough to modify the distribution of pulsars between the two kick velocity  modes significantly. 
Currently, it is very difficult to prove the importance of this effect on our results directly, as the number of pulsars with well-established obliquity is not high enough to obtain statistically sound results in our approach. 
However, we checked whether there is any difference in the alignment properties between pulsars from the two modes. 

As radio waves can be emitted in the so-called orthogonal polarization
mode, there is a 90-degree ambiguity in the position of the spin axis, see, e.g.,
\cite{2007ApJ...664..443R}. In this case, the emission is polarized perpendicular to the magnetic field lines. For example, for the Vela pulsar, the spin direction is determined independently using X-ray data, and it is possible to demonstrate
that the polarization measurements define the direction perpendicular to
the spin axis orientation \citep{2005MNRAS.364.1397J}. 
It is sometimes assumed that the angle between the
velocity vector and the direction derived from the polarization measurements
is $< 45^\circ$ by definition. 
I.e., if the measured angle $\mathrm{PA}_\mathrm{meas}=\left|\mathrm{PA}_\mathrm{V}-\mathrm{PA}_0 \right|$, under the assumption of the parallel
polarization mode appears to be $\mathrm{PA}_\mathrm{meas} > 45^\circ$, then it is assumed that the
emission is in the orthogonal mode and the spin axis orientation corresponds
to $90^\circ-\mathrm{PA}_\mathrm{meas}$, see \cite{2007ApJ...660.1357N,2012MNRAS.423.2736N}.
Here $\mathrm{PA}_\mathrm{V}$ is the position angle of a pulsar's proper motion vector on the sky, and $\mathrm{PA}_0$ is the polarization position angle obtained by fitting a rotating vector model \citep{1969ApL.....3..225R}, which is used as a proxy for the orientation of the rotation axis.

\begin{figure}[h]
\centering
\includegraphics[width=0.45\textwidth]{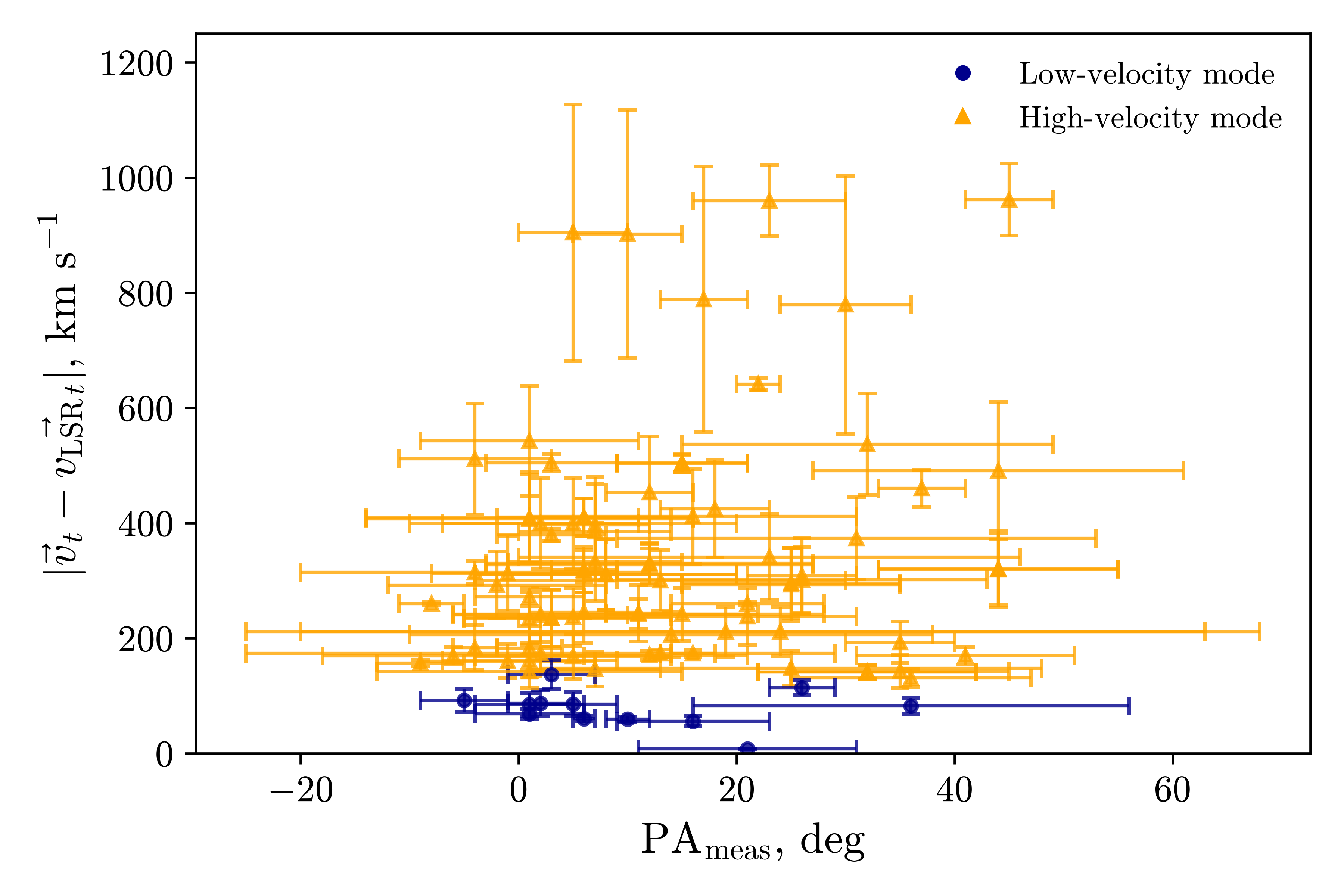}
\caption{Spin-velocity alignment angle $\mathrm{PA}_\mathrm{meas}=\left|\mathrm{PA}_\mathrm{V}-\mathrm{PA}_0\right|$ vs. transverse velocity relative to the transverse velocity component of LSR velocity. Pulsars from the two modes of the kick velocity are shown with different symbols and colors as in the previous figures. Error bars for $|\vec{v}_t - \vec{v_\text{LSR}}_t|$ represent a 1$\sigma$ interval of the quantity for the sample of trajectories of a pulsar.
}
\label{fig:PPA}
\end{figure}

We checked if any pulsars with reported $\mathrm{PA}_\mathrm{meas}$ are found in our sample. 
In order to verify whether spin-velocity correlation is observed, for each trajectory of a pulsar we evaluate the transverse component of the pulsar's velocity relative to the velocity of LSR. 
This value is chosen instead of the kick velocity because it is independent of our assumption of velocity---line-of-sight angle distribution. 
Then we source $\mathrm{PA}_\mathrm{meas}$ as $\left|\mathrm{PA}_\mathrm{V}-\mathrm{PA}_0\right|$ from \cite{2007ApJ...660.1357N}, \cite{2015MNRAS.453.4485F}, \cite{2007MNRAS.381.1625J}, and \cite{2015ApJ...804..112R}. Additionally, we flip the value of $\mathrm{PA}_\mathrm{meas}$ to $90^\circ-\mathrm{PA}_\mathrm{meas}$ if $\mathrm{PA}_\mathrm{meas}>45^\circ$.
The results are presented in Fig.~\ref{fig:PPA}. 
In total, we show $\mathrm{PA}_\mathrm{meas}$ for 10 low-velocity pulsars and 56 high-velocity pulsars.
Visibly, there is no correlation between the alignment angle and the kick velocity mode. 
Also, we do not see any correlation between the angle and the transverse velocity value for any of the modes or for the sample in general.
It is worth noting that most of the low-velocity pulsars with $\mathrm{PA}_\mathrm{meas}\approx 0^\circ\pm 10^\circ$ come from the sample presented by \cite{2015ApJ...804..112R}.  While all low-velocity pulsars with $|\mathrm{PA}_\mathrm{meas}| \gtrsim 15^\circ$ come from other samples we used. 

\section{Discussion}\label{sect:discussion}

In this section, we discuss the interpretation and implications of our study, as well as some assumptions and uncertainties.
First, we provide a rationale for several assumptions we adopt for our model and the limitations of our study.
Then, we discuss in detail the results of our analysis and provide possible explanations for the differences between the two distribution modes.

\subsection{Model assumptions}

Our model relies on several assumptions, specifically, on constraints on the radial velocities and ages of the pulsars.

Derivation of 3D velocities of radio pulsars is sensitive to the choice of their radial velocities, which cannot be measured directly. Various assumptions can be used here.
E.g., in their study,~\cite{2024MNRAS.527.1101G} use a method that differs from ours in two ways: sampling the angle between the line-of-sight and 3D velocity vector and the fact that they consider two different rest frames, in which a pulsar's possible velocity vectors are isotropic.

\subsubsection{Velocity---line-of-sight angle sampling}

In our study, in Sect.~\ref{sect:model} we sample the angles between the line of sight velocity vectors from a uniform distribution over the central 98\% interval of $(0, \pi)$, i.e. $\theta$ is sampled from $\text{Uniform}(0.01\pi, \, 0.99\pi)$.
This differs from the original technique used by~\cite{2024MNRAS.527.1101G}, who sampled $\cos{\theta}$ from $\text{Uniform}(-1,\,1)$ (i.e. they sampled $\theta$ from a distribution with probability density $f(\theta)=\frac{1}{2}\sin\theta$) following the distribution of the polar angle $\theta$ of isotropically oriented unit vectors. 
To test the difference between the two approaches we run the procedure of our main model, except we sample $\cos\theta$ from $\text{Uniform}(-0.98,\,0.98)$.

This model produces a slightly larger fraction of low-velocity pulsars: $\sim25\%$, versus $\sim23\%$ obtained using our main model. 
Additionally, all the pulsars that belong to the low-velocity mode in our main model also belong to the low-velocity mode for this simulation, i.e. sampling $\cos{\theta}\sim\text{Uniform}(-0.98,\,0.98)$ just increases the number of low-velocity pulsars.

This can be explained by the fact that sampling $\theta$ from $\text{Uniform}(0.01\pi, \, 0.99\pi)$ yields a larger number of angles that are closer to either 0 or $\pi$ than uniformly sampling $\cos\theta$, meaning it produces larger $\left|\cot\theta\right|$. 
Although our model's angle sampling method differs from~\cite{2024MNRAS.527.1101G}, our results appear stable with respect to the choice of line-of-sight---velocity angle sampling method.

\subsubsection{Galactocentric and local-standard-of-rest isotropy}
\label{sect:discussion:GC}

Additionally, \cite{2024MNRAS.527.1101G} employ two approaches regarding the isotropy of binary NSs' velocity vectors: isotropy with respect to the Galactocentric frame (GC) and to the pulsar's local standard of rest (LSR).

In our calculations presented above, we used the latter approach. 
To test the difference in these assumptions, we also performed calculations under the assumption of isotropy with respect to the GC frame, leaving all other ingredients of the model the same. 
That is, we sample pulsars' radial velocities with an equation similar to Eq.~\ref{eq:infer_vr}:
\begin{equation}
    v_r=\left|\vec{v}_t\right|\hspace{.3mm}\cot\theta.
    \label{eq:infer_vr_GC}
\end{equation}
As a result,
for the GC isotropy, only 18 pulsars ($\sim9\%$) belong to the low-velocity mode. This number is significantly less than the $\sim23\%$ obtained for the LSR isotropy.
All of the low-velocity pulsars in the GC case belong to the low-velocity mode in the LSR case as well, i.e., the GC isotropy assumption just reduces the number of objects in the low-velocity mode. 

This difference arises because the GC isotropy assumption mostly affects pulsars with small kicks.
Following the assumption that the kick velocity is isotropic with respect to the LSR at its birthplace, full Galactocentric 3D velocities at birth for pulsars with small kicks are close to the velocity of the LSR.
This means that the transverse component of the peculiar velocity $\left|\vec{v}_t-\vec{v}_{\text{LSR},\,t}\right|$ is small, whereas $\left|\vec{v_t}\right|$ itself will be closer to the projection of the circular rotation velocity in the Milky Way as we already introduced a correction for the Galactic motion of the Sun. 
After inferring the radial velocity based on these two assumptions, the GC isotropy assumption will produce a value of $v_r$ much larger than that produced by the LSR isotropy assumption. 
This does not significantly affect larger kicks; however, smaller kicks are overestimated under the GC isotropy assumption.
This, in turn, results in a smaller fraction of pulsars in the low-velocity mode,
as it was also shown by \cite{2024MNRAS.527.1101G} who claimed that GC isotropy tends to underestimate low velocities and overestimates the corresponding kick magnitude.


\subsubsection{Characteristic age}

An important assumption in our modeling is related to the characteristic age $\tau_c$. 
The spin-down ages of the pulsars are not an accurate representation of their true ages, as in their derivation, it is assumed that the initial spin period is much shorter than the present one and the evolution proceeds with a constant braking index, typically taken as $n=3$.  
Initial spin periods can be comparable with the observed values, especially for young NSs, e.g. \cite{2012Ap&SS.341..457P}. Braking indices can change during the lifetime of a pulsar, e.g., due to decaying \citep{2013MNRAS.432..967I} or re-emerging \citep{2016MNRAS.462.3689I} magnetic fields.
This results in a difference between the kinematic and characteristic ages, e.g. \cite{2013MNRAS.430.2281N}. However, characteristic age is the only age estimate that can be obtained for many pulsars in a clear and uniform way. Thus, we prefer to use this value in our modeling to determine one of the possible birthplaces for each pulsar trajectory. 

Nevertheless, to test the extent to which the choice of the integration limits for trajectories impacts our results, we halve the assumed ages of the pulsars and repeat the steps of our analysis.
For this model, we use $0.75\tau_c$ for the time integration limit of pulsars' trajectories, and similarly to the $t=-\tau_c$ birthplaces of our main model, we collect kick velocities at the points of the trajectories where $t=-0.5\tau_c$.
Our results do not change significantly: only five pulsars ($<3\%$) are categorized differently in comparison to our main model.

\subsection{Magnetic fields}

\begin{figure}[h]
\centering
\includegraphics[width=0.45\textwidth]{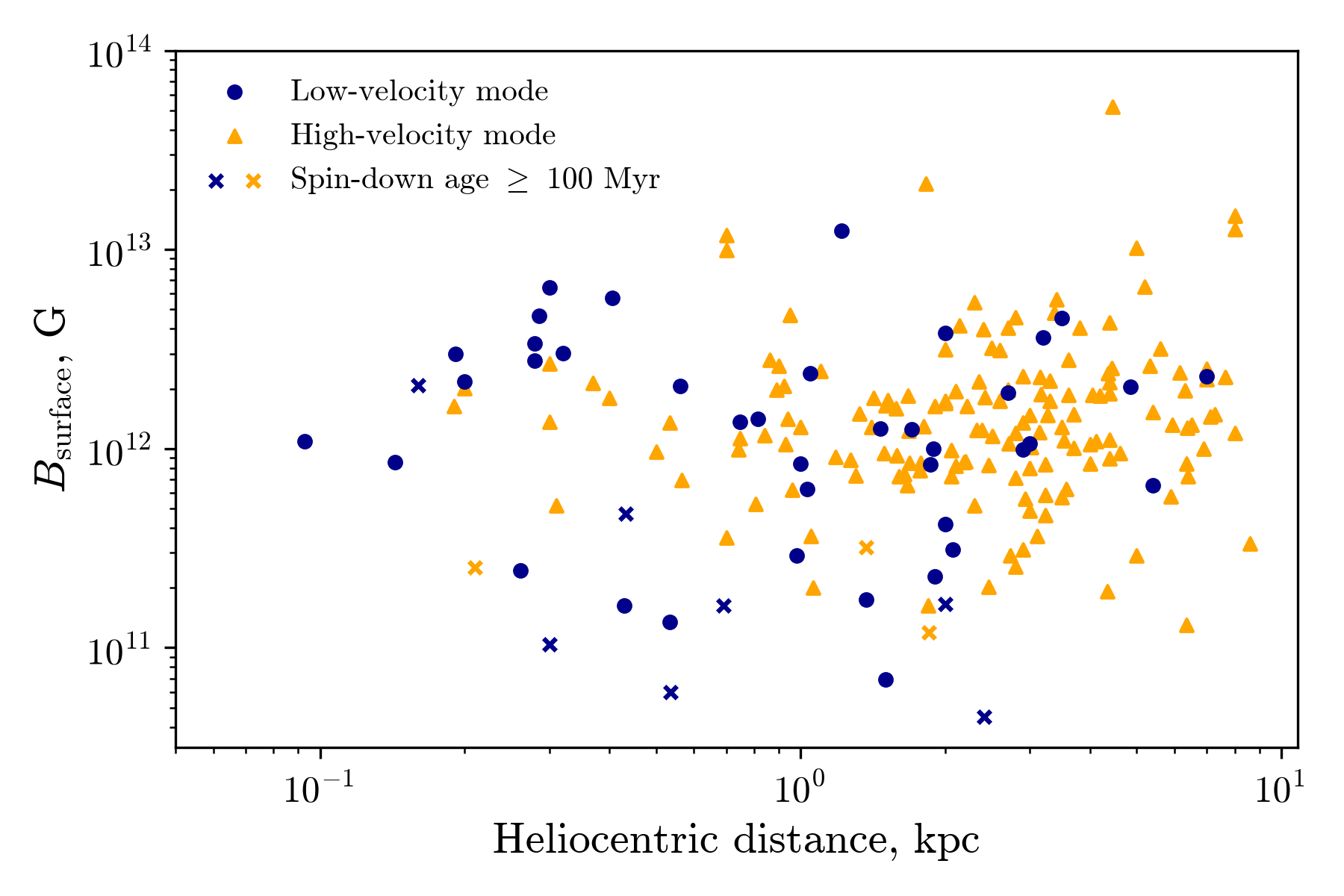}
\caption{Effective surface dipolar magnetic field of pulsars, plotted against their Galactocentric distances. 
Low-velocity and high-velocity distribution modes are represented by dark blue and orange, respectively.
Pulsars with large spin-down ages ($\ge 100$~Myr) are represented by crosses.
Both axes are logarithmic.
}\label{fig:B_surf_D}
\end{figure}

Previously, the correlation between velocity and magnetic field of radio pulsars was not reported, see e.g., \cite{1999A&A...351..195D}. However, theoretical proposals, based on evolution in binary systems, for such a correlation have been put forward; see, e.g., \cite{1989ApJ...342..917B} as an early example.
In our study, we found that the magnetic field distribution differs between the two kick velocity modes. The origin of this difference is uncertain. In part, it can be attributed to selection effects. E.g., the average distance for high-velocity pulsars is larger. Thus, they might have higher intrinsic luminosities. This, in its turn, can require higher rotational energy losses, and so---higher magnetic fields. However, note that at the high end of the field distribution, the pulsars from the two modes nearly coincide. 
In addition, as it is visible in Fig.~\ref{fig:B_surf_D}, low-velocity pulsars dominate among low-field objects at all distances.

Pulsars in the low-velocity mode are older (on average) than those in the high-velocity mode. 
So, the excess of low-field pulsars in the low-velocity mode can be attributed to the magnetic field decay as pulsars with larger spin-down ages have smaller fields, see Fig.~\ref{fig:B_surf_t}. However, various studies, e.g. \cite{2006ApJ...643..332F}, do not demonstrate significant field evolution in normal pulsars (see, however, \cite{2024ApJ...972...78S} and references therein). Just a mild decay by a factor $\sim 2$ at ages $\lesssim 10^6$~yrs was proposed in \cite{2014MNRAS.444.1066I}. This can hardly explain the observed difference in the field distribution. 

In Fig.~\ref{fig:B_surf_t}, it is seen that the main difference between the magnetic field distributions for the two modes appears at ages $\gtrsim \text{few} \times10^6$~yrs. The difference in the characteristic ages between the two distributions (see Fig.~\ref{fig:CDFs}) reaches nearly an order of magnitude for $\tau_c\gtrsim 10^7$~yrs. As the shift between the two magnetic field distributions (see Fig.~\ref{fig:BSURF_CDF}) is less than an order of magnitude, it potentially can be attributed to the magnetic field decay with a characteristic time scale $\gtrsim 10^7$~yrs. Similar time scales are discussed in application to various individual objects, see e.g., \cite{2020Univ....6...63W} and references therein.
On the other hand, if the magnetic field is decaying, then braking indices are expected to have values $\gg3$. In Fig.~\ref{fig:n}, we see that the difference between pulsars from the low- and high-velocity modes appears not at $n\gg3$ but at $n<0$. Of course, measured braking indices can have peculiar values not because of the magnetic field decay or re-emergence but due to glitches and noisy behavior of radio pulsars. Thus, such an argument based on poorly determined values of $n$ is a weak one. 
Still, detailed population synthesis modeling is welcome, also because the characteristic age is a poor parameter for many estimates due to crucial assumptions made in its definition (no field decay, no magnetic inclination evolution, and a small initial period). 

\begin{figure}[h]
\centering
\includegraphics[width=0.4\textwidth]{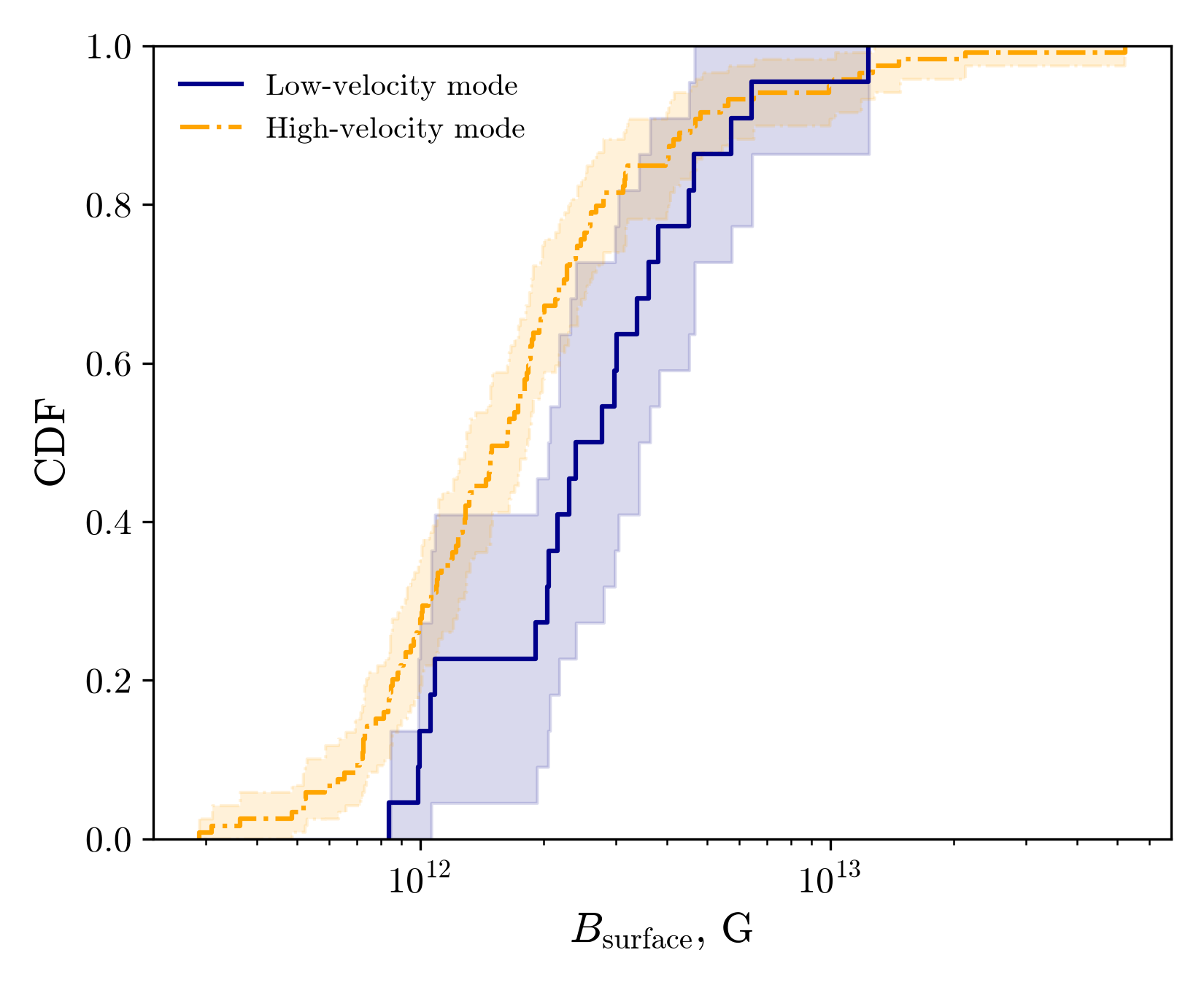}
\caption{Cumulative distribution functions of the effective surface dipolar magnetic field of pulsars with $\tau_c < 10{\rm \space Myr}$. 
Low-velocity and high-velocity distribution modes are represented by dark blue and orange, respectively. 
Shaded areas represent a bootstrapped 95\% confidence interval. 
The horizontal axis is on a logarithmic scale. The CDFs are shown for the full pulsar sample.
}\label{fig:BSURF_CDF_young}
\end{figure}

Another uncertainty is related to the ages of radio pulsars in our sample. As we noticed above, the characteristic age is often a poor measure of the true age. Fig.~\ref{fig:B_surf_t} illustrates that if we introduce a cut eliminating pulsars with larger $\tau_c$, then due to poor statistics of low-velocity pulsars, results become ambiguous. E.g., if we plot the CDF for magnetic field in the low- and high-velocity modes only for pulsars with $\tau_c<10$~Myr (see Fig.~\ref{fig:BSURF_CDF_young}), then we see a significant difference between the two distributions ($p$-value $\sim0.002$), but the relation is reversed in comparison to Fig.~\ref{fig:BSURF_CDF}. I.e., the low-field part is dominated by high-velocity pulsars. A look at Fig.~\ref{fig:B_surf_t} suggests that this is due to a small number of low-velocity pulsars with $B\sim$a few$\times 10^{11}$~G and $\tau_c\sim$a few$\times 10^6$~yrs. This behaviour can be considered as an argument against the physical origin of the difference observed in Fig.~\ref{fig:BSURF_CDF}. With the present-day statistics and poorly known ages of pulsars, it is premature to make solid conclusions.

If the bimodal structure of the kick velocity is due to some differences in the physics of SN explosions, it is tempting to suggest that the difference between the two field distributions has a similar origin. The three main families of NS kick models are the following (see an early review in \citealt{2003ASPC..302..307L}): hydrodynamical \citep{2010ApJ...725L.106W,  2024ApJ...964L..16B}, asymmetric neutrino emission \citep{2019ApJ...880L..28N, 2024Ap&SS.369...80J}, and magnetorotational \cite{2024PhRvD.110h3025K, 2023MNRAS.522.6070P}. 
Potentially, in all three, it is possible to speculate about a possible correlation between the (external dipolar) magnetic field of the NS and the kick. However, a detailed analysis of this issue is beyond the scope of our study.

\begin{figure}[h]
\centering
\includegraphics[width=0.45\textwidth]{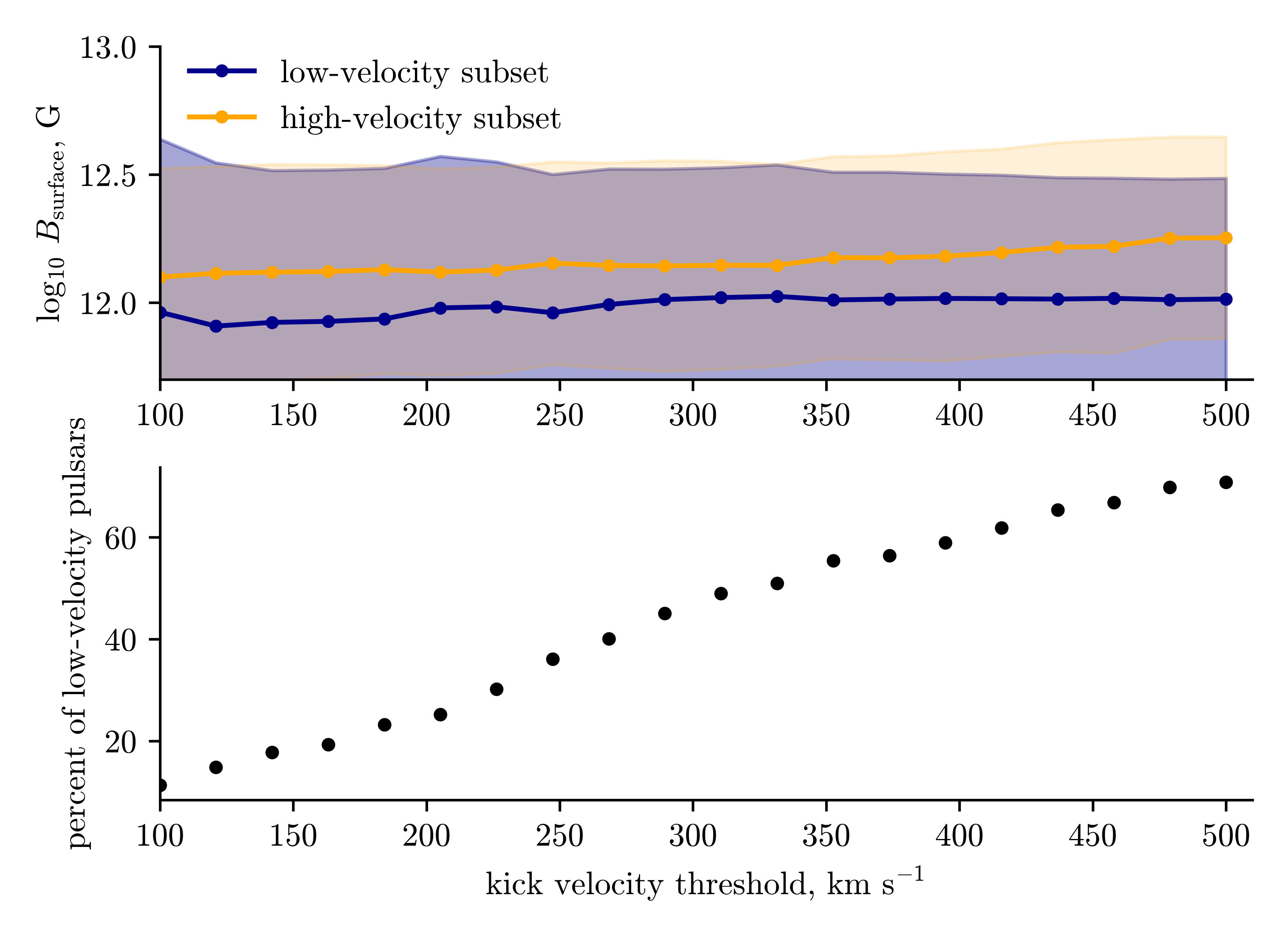}
\caption{
Logarithm of magnetic fields of pulsars with lower and higher kick velocities.
The upper panel shows the average logarithm of the effective surface magnetic fields of the two subsets of pulsars, plotted against the subset threshold velocity.
The low-velocity subset and high-velocity subset are shown in dark blue and orange, respectively. 
Areas shaded blue and orange represent pulsars' magnetic fields within one standard deviation from the average logarithm at each kick threshold value.
The lower panel shows the percentage of pulsars assigned to the low-velocity subset for each kick threshold value.}
\label{fig:B_splits}
\end{figure}

Additionally, we provide a simple test to see whether the kick velocity bimodality manifests itself in the magnetic field data for different values of the threshold velocity.
For each pulsar, we compute the median of its possible kick velocity values obtained for various trajectories for both Galactic crossing birthplaces and the $t=-\tau_c$ birthplaces. 
Then, for a set of 20 threshold values from 100 \kms to 500 \kms, we form two subsets of pulsars: pulsars with a median kick smaller and larger than the threshold value.
We then evaluate the average and the standard deviation of the magnetic field of the two pulsar subsamples for each threshold value. The results are shown in Figure~\ref{fig:B_splits}.

For all kick velocity threshold values, average magnetic fields of the high-velocity subset are larger than those of the low-velocity subset.
Despite minor variations for different threshold values, we do not observe any sign of a clear trend in the average magnetic field behavior.
Thus, we conclude that the magnetic field distribution is not sensitive to the threshold value of a bimodal kick distribution. 



\section{Conclusions}
\label{sect:conclusions}
    
 In this study, we classified 202 normal radio pulsars according to their kick velocities into two groups. These two groups correspond to the low- and high-velocity modes of the kick velocity distribution proposed in \cite{2020MNRAS.494.3663I}. Approximately 23\% of the pulsars from our sample belong to the low-velocity mode, in correspondence with \cite{2020MNRAS.494.3663I}. 

 Our main goal was to compare the distributions of several pulsar parameters across the two modes of the kick velocity distribution. We found that NSs from the low-velocity group on average have lower magnetic fields. The effect is mainly pronounced for $B\lesssim10^{12}$~G.
The origin of this difference is uncertain. 
Although it is tempting to suggest that it can be partly related to the supernova explosion mechanism, which is responsible for the bimodal kick velocity distribution, our analysis suggests that it can be due to selection effects and/or low statistics. Thus, we do not see any robust dichotomy in radio pulsar parameters correlated with the proposed hypothetical kick bimodality.

\section*{Acknowledgements}

We are grateful to Drs. Anton Biryukov and Andrei Igoshev for discussions and comments. We thank the referee for useful comments, which helped to improve the manuscript.
The study was conducted under the state assignment of Lomonosov Moscow State University.

\section*{Author contribution}

We declare that this manuscript is original and has not been published elsewhere. Both authors contributed equally to this work.

\section*{Data availability}

The data used in this study were obtained from the Australia Telescope National Facility (ATNF) Pulsar Catalogue v2.7.0, available at https://www.atnf.csiro.au/research/pulsar/psrcat/.
The list of pulsars belonging to each of the kick velocity modes is available on request.

\section*{Funding}

We declare that no funds, grants, or other support were received during the preparation of this manuscript.

\section*{Declarations}

\subsection*{Competing interests}

We declare no competing interests.

\subsection*{Ethics declaration}

Not applicable.

\bibliographystyle{mnras}
\bibliography{kick_bimodal}

\end{document}